\documentclass[12pt, draftclsnofoot, onecolumn]{IEEEtran}
\normalsize

\ifCLASSINFOpdf
\else
\fi

\usepackage[font=footnotesize]{subfig}
\usepackage{array}
\usepackage{fixltx2e}
\usepackage[cmex10]{amsmath}
\usepackage[font=footnotesize, margin=0.0cm]{caption}
\usepackage{graphicx}
\usepackage{amsmath}
\usepackage{amsthm, amssymb}
\usepackage[overload]{empheq}
\usepackage{commath}
\usepackage{caption}
\usepackage{multirow}
\usepackage{algorithm}
\usepackage{algorithmic}
\newcommand{\RN}[1]{%
  \textup{\uppercase\expandafter{\romannumeral#1}}%
  }
\usepackage{mathtools}

\DeclareMathAlphabet{\mathpzc}{OT1}{pzc}{m}{it}
\usepackage{cite}
\usepackage{color}
\usepackage[utf8]{inputenc}
\usepackage[english]{babel}

\usepackage{url}
\usepackage{setspace}
\usepackage{changebar}

\usepackage{url}

\graphicspath{{../fig/}}   % Location of your graphics files

\begin{document}
%
% paper title
% can use linebreaks \\ within to get better formatting as desired
\title{Resource Allocation for Intelligent Reflecting Surface Aided Wireless Powered Mobile Edge Computing in OFDM Systems}
\author{Tong Bai, \IEEEmembership{Member,~IEEE},
		Cunhua Pan, \IEEEmembership{Member,~IEEE},\\
		Hong Ren, \IEEEmembership{Member,~IEEE},
		Yansha Deng, \IEEEmembership{Member,~IEEE}, \\
		Maged Elkashlan, \IEEEmembership{Member,~IEEE},
		and Arumugam Nallanathan, \IEEEmembership{Fellow,~IEEE}
\thanks{T. Bai, C. Pan, H. Ren, M. Elkashlan and A. Nallanathan are with the School of Electronic Engineering and Computer Science, Queen Mary University of London, London E1 4NS, U.K. (e-mail: t.bai@qmul.ac.uk, c.pan@qmul.ac.uk, h.ren@qmul.ac.uk, maged.elkashlan@qmul.ac.uk, a.nallanathan@qmul.ac.uk). Y. Deng is with the Department of Engineering, King’s College London, London, WC2R 2LS, U.K. (e-mail: yansha.deng@kcl.ac.uk).}% <-this % stops a space
}

% make the title area
\maketitle
\IEEEpeerreviewmaketitle

\vspace{-1.6cm}
\begin{abstract}
Wireless powered mobile edge computing (WP-MEC) has been recognized as a promising technique to provide both enhanced computational capability and sustainable energy supply to massive low-power wireless devices. However, its energy consumption becomes substantial, when the transmission link used for wireless energy transfer (WET) and for computation offloading is hostile. To mitigate this hindrance, we propose to employ the emerging technique of intelligent reflecting surface (IRS) in WP-MEC systems, which is capable of providing an additional link both for WET and for computation offloading.
Specifically, we consider a multi-user scenario where both the WET and the computation offloading are based on orthogonal frequency-division multiplexing (OFDM) systems.
Built on this model, an innovative framework is developed to minimize the energy consumption of the IRS-aided WP-MEC network, by optimizing the power allocation of the WET signals, the local computing frequencies of wireless devices, both the sub-band-device association and the power allocation used for computation offloading, as well as the IRS reflection coefficients. The major challenges of this optimization lie in the strong coupling between the settings of WET and of computing as well as the unit-modules constraint on IRS reflection coefficients. To tackle these issues, the technique of alternative optimization is invoked for decoupling the WET and computing designs, while two sets of locally optimal IRS reflection coefficients are provided for WET and for computation offloading separately relying on the successive convex approximation method. The numerical results demonstrate that our proposed scheme is capable of monumentally outperforming the conventional WP-MEC network without IRSs. Quantitatively, about $80\%$ energy consumption reduction is attained over the conventional MEC system in a single cell, where $3$ wireless devices are served via $16$ sub-bands, with the aid of an IRS comprising of $50$ elements.
\end{abstract}

%\begin{IEEEkeywords}
%Intelligent reflecting surface, mobile edge computing, latency minimization.
%\end{IEEEkeywords}

\section{Introduction}
\label{sec:Introduction}

\subsection{Motivation and Scope}

In the Internet-of-Things (IoT) era, a myriad of heterogeneous devices are envisioned to be interconnected \cite{8879484}. However, due to the stringent constraints both on device sizes and on manufacturing cost, many of them have to be equipped with either life-limited batteries or low-performance processors. Consequently, if only relying on their local computing, these resource-constrained devices are incapable of accommodating the applications that require sustainable and low-latency computation, e.g. wireless body area networks \cite{8703166} and environment monitoring \cite{saffari2017rf}.
Fortunately, wireless powered mobile edge computing (WP-MEC)\cite{you2016energyMEC,wang2018joint,bi2018computation,
feng2019computation,wu2019online,huang2019deep,hu2018wireless,
ji2018energy,zhou2018computation,liu2019uav}, which incorporates radio frequency (RF) based wireless energy transmission (WET) \cite{bi2015wireless,huang2015cutting,niyato2017wireless} and mobile edge computing (MEC) \cite{barbarossa2014communicating,shi2016edge}, constitutes a promising solution of this issue.
Specifically, at the time of writing, the commercial RF-based WET has already been capable of delivering $0.05~\rm{mW}$ to a distance of $12-14~\rm{m}$ \cite{bi2015wireless}, which is sufficient to charge many low-power devices, whilst the MEC technique may provide the cloud-like computing service at the edge of mobile networks \cite{shi2016edge}.
In WP-MEC systems, hybrid access points (HAP) associated with edge computing nodes are deployed in the proximity of wireless devices, and the computation of these devices is typically realized in two phases, namely the WET phase and the computing phase. To elaborate, the batteries of the devices are replenished by harvesting WET signals from the HAP in the first phase, while in the computing phase, devices may decide whether to process their computational tasks locally or offload them to edge computing nodes via the HAP.

Given that these wireless devices are fully powered by WET in WP-MEC systems, the power consumption at HAPs becomes substantial, which inevitably increases the expenditure on energy consumption and may potentially saturate power rectifiers. At the time of writing, the existing research contributions that focus on reducing the power consumption mainly rely on the joint optimization of the WET and of computing \cite{wang2018joint}, as well as cooperative computation offloading \cite{hu2018wireless,ji2018energy}.
However, wireless devices are still suspicious to severe channel attenuation, which limits the performance of WP-MEC systems. To resolve this issue, we propose to deploy the emerging intelligent reflecting surfaces (IRS) \cite{wu2019towards,basar2019wireless,ntontin2019reconfigurable} in the vicinity of devices, for providing an additional transmission link both for WET and for computation offloading. Then, the power consumption can be beneficially reduced both for the downlink and for the uplink.
To elaborate, an IRS comprises of an IRS controller and a large number of low-cost passive reflection elements. Regulated by the IRS controller, each IRS reflection element may adapt both the amplitude and the phase of the incident signals reflected, for collaboratively modifying the signal propagation environment. The gain attained by IRSs is based on the combination of so-called the virtual array gain and the reflection-enabled beamforming gain \cite{wu2019towards}. More explicitly, the virtual array gain is achieved by combining the direct and IRS-reflected links, while the reflection-enabled beamforming gain is realized by proactively adjusting the reflection coefficients of the IRS elements. By combining these two types of gain together, IRSs are capable of reducing the power required both for WET and for computation offloading, thus improving the energy efficiency of WP-MEC systems. In this treatise, we aim for providing a holistic scheme to minimize the energy consumption of WP-MEC systems, relying on IRSs.

\subsection{Related Works}

The current state-of-the-art contributions are reviewed from the perspectives of WP-MEC and of IRS-aided networks, as follows.
\subsubsection{Wireless Powered Mobile Edge Computing}
This topic has attracted an increasing amount of research attention \cite{you2016energyMEC,wang2018joint,bi2018computation,
feng2019computation,wu2019online,huang2019deep,hu2018wireless,
ji2018energy,zhou2018computation,liu2019uav}.
Specifically, You \emph{et al.} firstly proposed the WP-MEC framework \cite{you2016energyMEC}, where the probability of successfully computing was maximized subject to the constraints both on energy harvesting and on latency.
The single-user system considered in this first trial limits its application in large-scale scenarios.
For eliminating this shortage, an energy-minimization algorithm was proposed for the multi-user scenario \cite{wang2018joint}, where the devices' computation offloading was realized by the time division multiple access (TDMA) technique.
Following this, Bi and Zhang maximized the weighted sum computation rate in a similar TDMA system \cite{bi2018computation}, while an orthogonal frequency division multiple access (OFDMA) based multi-user WP-MEC system was investigated in \cite{feng2019computation}.
A holistic online optimization algorithm was proposed for the WP-MEC in industrial IoT scenarios \cite{wu2019online}.
In the aforementioned works, the associated optimization is commonly realized with the aid of the alternative optimization (AO) method, because the pertinent optimization problems are usually not jointly convex. This inevitably imposes a delay on decision making. To mimic this issue, Huang \emph{et al.} proposed a deep reinforcement learning based algorithm for maximizing the computation rate of WP-MEC systems \cite{huang2019deep}, which may replace the aforementioned complicated optimization by a pre-trained look-up table.
Furthermore, as for the system where both near and far devices have to be served, the energy consumption at the HAP has to be vastly increased, because the farther device harvests less energy while a higher transmit power is required for its computation offloading.
Aimed for releasing this so-called ``doubly near-far" issue, the technique of user cooperation was revisited \cite{hu2018wireless,ji2018energy}.
%Apart from collaborating with terrestrial cellular systems as aforementioned, WP-MEC systems can also be mounted on the unmanned aerial vehicles (UAV) \cite{zhou2018computation,liu2019uav}, for exploiting their agility and on-demand deployment in dynamic networks.
At the time of writing, the WET and computation offloading in WP-MEC systems in the face of hostile communication environments has not been well addressed. Against this background, we aim for tackling this issue by invoking IRSs. Let us now continue by reviewing the relevant research contributions on IRSs as follows.

\subsubsection{IRS-Aided Networks}

In order to exploit the potential of IRSs, an upsurging number of research efforts have been devoted in its channel modeling \cite{tang2019wireless,ozdogan2019intelligent}, analyzing the impact of limited-resolution phase shifts \cite{han2019large,di2020practical}, channel estimation \cite{hu2019two,yang2019intelligent} as well as IRS reflection coefficient designs \cite{wu2019intelligent,wu2019beamforming,guo2019weighted,
zhou2020framework}.
Inspired by these impressive research contributions, the beneficial role of IRSs was evaluated in various application scenarios \cite{pan2019intelligent,yang2019irs,dong2020secure,hong2020artificial,
pan2019intelligent_swipt,zheng2020intelligent,bai2019latency}.
Specifically, an IRS was employed in multi-cell communications systems for mitigate the severe inter-cell interference \cite{pan2019intelligent}, where an IRS comprising of $100$ reflection elements was shown to be capable of doubling the sum rate of the multi-cell system.
Yang \emph{et al.} investigated an IRS-enhanced OFDMA system \cite{yang2019irs}, whose common rate was improved from around $2.75~\rm{bps/Hz}$ to $4.4~\rm{bps/Hz}$, with the aid of a $50$-element IRS.
Apart from assisting the aforementioned throughput maximization in the conventional communications scenario, a sophisticated design of IRSs may also eminently upgrade the performance of diverse emerging wireless networks, e.g. protecting data transmission security \cite{dong2020secure,hong2020artificial}, assisting simultaneous wireless information and power transfer (SWIPT) \cite{pan2019intelligent_swipt}, enhancing the user cooperation in wireless powered communications networks \cite{zheng2020intelligent}, as well as reducing the latency in IRS-aided MEC systems \cite{bai2019latency}. These impressive research contributions inspire us to exploit the beneficial role of IRSs in this momentous WP-MEC scenario.

\subsection{Novelty and Contributions}

In this paper, an innovative IRS-aided WP-MEC framework is proposed, where we consider orthogonal frequency-division multiplexing (OFDM) systems for its WET and devices' computation offloading. Under this framework, a joint WET and computing design is conceived for minimizing its energy consumption, by optimizing the power allocation of the WET signals over OFDM sub-bands, the local computing frequencies of wireless devices, both the sub-band-device association and the power allocation used for computation offloading, as well as the pertinent IRS reflection coefficient design. Let us now detail our contributions as follows.
\begin{itemize}
\item \emph{Energy minimization problem formulation for the new IRS-aided WP-MEC design:} In order to reduce the energy consumption of WP-MEC systems, we employ an IRS in WP-MEC systems and formulate a pertinent energy minimization problem. Owing to the coupling effects between the designs of WET and of computing, it is difficult to find its globally optimal solution. Alternatively, we provide an alternative optimization (AO) based solution to approach a locally optimal solution, by iteratively optimizing settings of WET and of computing.
\item \emph{WET design:} The WET setting is realized by alternatively optimizing the power allocation of energy-carrying signals over OFDM sub-bands and the IRS reflection coefficients. Specifically, given a set of fixed IRS reflection coefficients, the power allocation problem can be simplified to be a linear programming problem, which can be efficiently solved by the existing optimization software. Given a fixed power allocation, the IRS reflection coefficient design becomes a feasibility-check problem, the solution of which is incapable of ensuring a rapid convergence. To tackle this issue, we reformulate the problem by introducing a number of auxiliary variables, and provide a locally optimal design of IRS reflection coefficients, with the aid of several steps of mathematical manipulations and of the successive convex approximation (SCA) method.
\item \emph{Computing design:}
The settings at the computing phase are specified by alternatively optimizing the joint sub-band-device association for and the power allocation for devices' computation offloading, IRS reflection coefficients at the computing phase as well as the local computing frequencies. Specifically, as verified by \cite{seong2006optimal}, the duality gap vanishes when the number of sub-bands exceeds $8$. Hence, we provide a near-optimal solution for the joint sub-band-device association and power allocation problem, relying on the Lagrangian duality method. The IRS reflection coefficients are designed using the similar approach devised for that in the WET phase. Finally, our analysis reveals that the optimal local computing frequencies can be obtained by selecting their maximum allowable values.
\item \emph{Numerical validations:} Our numerical results validates the benefits of employing IRSs in WP-MEC systems, and quantify the energy consumption of our proposed framework in diverse simulation environments, together with two benchmark schemes.
\end{itemize}

The rest of the paper is organized as follows. In Section~\RN{2}, we describe the system model and formulate the pertinent problem. A solution of this problem is provided in Section~\RN{3}. The numerical results are presented in Section~\RN{4}. Finally, our conclusions are drawn in Section~\RN{5}.

\begin{figure}[h!]\center
\includegraphics[width= 0.5 \textwidth]{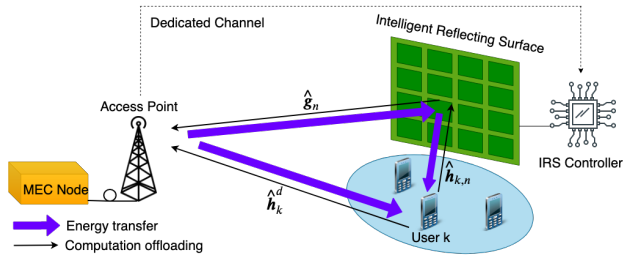}
\caption{An illustration of our IRS-aided WP-MEC system, where $K$ single-antenna devices are served by a mobile edge computing node via a single-antenna hybrid access point, with the aid of an $N$-element IRS.}
\label{fig:system_model}
%\hrulefill
\end{figure}

\section{System Model and Problem Formulation}
\label{sec:System Model}

\begin{figure}[h!]\center
\includegraphics[width= 0.4 \textwidth]{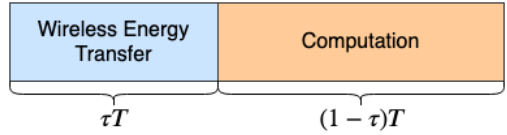}
\caption{An illustration of the harvest-then-offloading protocol, where $\tau T$ and $(1-\tau)T$ refer to the duration of the WET and the computing phases, respectively.}
\label{fig:harvest_then_offloading}
\end{figure}

As illustrated in Fig.~\ref{fig:system_model}, we consider an OFDM-based WP-MEC system, where $K$ single-antenna devices are served by a single-antenna hybrid access point (HAP) associated with an edge computing node through $M$ equally-divided OFDM sub-bands.
Similar to the assumption in \cite{wang2018joint,bi2018computation,feng2019computation}, we assume that these devices do not have any embedded energy supply available, but are equipped with energy storage devices, e.g. rechargeable batteries or super-capacitors, for storing the energy harvested from RF signals. As shown in Fig.~\ref{fig:harvest_then_offloading}, relying on the so-called ``harvest-then-computing" mechanism \cite{wang2018joint}, the system operates in a two-phase manner in each time block. Specifically, during the WET phase, the HAP broadcasts energy-carrying signals to all $K$ devices for replenishing their batteries, while these $K$ devices process their computing tasks both by local computing and by computation offloading during the computing phase.
%Note that each device may harvest the energy of the RF signals transmitted over all $M$ sub-bands during the WET phase, while the devices are accessed to the system using OFDMA during computation offloading for avoid the co-channel interference.
We denote the duration of each time block by $T$, which is chosen to be no larger than the tolerant latency of MEC applications. The duration of the WET and of the computing phases are set as $\tau T$ and $(1-\tau)T$, respectively.
Furthermore, to assist the WET and the devices' computation offloading in this WP-MEC system, we place an IRS comprising of $N$ reflection elements in the proximity of devices.
The reflection coefficients of these IRS reflection elements are controlled by an IRS controller in a real-time manner, based on the optimization results provided by the HAP.

Let us continue by elaborating on the equivalent baseband time-domain channel as follows.
We denote the equivalent baseband time-domain channel of the direct link between the $k$-th device and the HAP, the equivalent baseband time-domain channel between the $n$-th IRS element and the HAP, and the equivalent baseband time-domain channel between the $k$-th device and the $n$-th IRS element by $\hat{\pmb{h}}^d_k \in \mathbb{C}^{L^d_{k}\times 1}$, $\hat{\pmb{g}}_n \in \mathbb{C}^{L_1\times 1}$ and $\hat{\pmb{r}}_{k,n} \in \mathbb{C}^{L_{2,k}\times 1}$, respectively, where $L^d_{k}$, $L_1$ and $L_{2,k}$ represent the respective number of delay spread taps. Without loss of generality, we assume that the above channels remain approximately constant over each time block. Furthermore, the channels are assumed to be reciprocal for the downlink and the uplink.

As for the IRS, we denote the phase shift vector of and the amplitude response of the IRS reflection elements by $\pmb{\theta} = [\theta_1, \theta_2, \ldots, \theta_N]^T$ and $\pmb{\beta} = [\beta_1, \beta_2, \ldots, \beta_N]^T$, respectively, where we have $\theta_n \in [0,2\pi)$ and $\beta_n \in [0,1]$. Then, the corresponding reflection coefficients of the IRS are given by $\pmb{\Theta} = [\Theta_1, \Theta_2, \ldots, \Theta_N]^T =  [\beta_1 e^{j\theta_1}, \beta_2 e^{j\theta_2}, \ldots, \beta_N e^{j\theta_N}]^T$, where $j$ represents the imaginary unit and we have $|\Theta_n| \leq 1$ for $\forall n \in \mathcal{N}$.
The baseband equivalent time-domain channel of the reflection link is the convolution of the device-IRS channel, of the IRS reflection response, and of the IRS-HAP channel. Specifically, the baseband equivalent time-domain channel reflected by the $n$-th IRS element is formulated as $\hat{\pmb{h}}^r_{k,n} = \hat{\pmb{r}}_{k,n} \ast \Theta_n \ast \hat{\pmb{g}}_n = \Theta_n \hat{\pmb{r}}_{k,n} \ast \hat{\pmb{g}}_n$. Here, we have $\hat{\pmb{h}}^r_{k,n} \in \mathbb{C}^{L^r_k \times 1}$ and $L^r_k = L_1 + L_{2,k} - 1$, which denotes the number of delay spread taps of the reflection channel. Furthermore, we denote the time-domain zero-padded concatenated device-IRS-HAP channel between the $k$-th device and the HAP via the $n$-th IRS element by $\pmb{v}_{k,n} = [(\hat{\pmb{r}}_{k,n} \ast \hat{\pmb{g}}_n)^T,0,\ldots,0]^T \in \mathbb{C}^{M\times 1}$. Upon denoting $\pmb{V}_{k} = [\pmb{v}_{k,1}, \ldots, \pmb{v}_{k,N}] \in \mathbb{C}^{M\times N}$, we formulate the composite device-IRS-HAP channel between the $k$-th device and the HAP as $\pmb{h}^r_{k} = \pmb{V}_{k} \pmb{\Theta} $. Similarly, we use $\pmb{h}^d_{k} = [(\hat{\pmb{h}}^d_{k})^T,0,\ldots,0]^T \in \mathbb{C}^{M\times 1}$ to represent the zero-padded time-domain channel of the direct device-HAP link. To this end, we may readily arrive at the superposed channel impulse response (CIR) for the $k$-th device, formulated as
\begin{IEEEeqnarray}{rCl}
\pmb{h}_k = \pmb{h}^d_k + \pmb{h}^r_k = \pmb{h}^d_k + \pmb{V}_k \pmb{\Theta}, \quad \forall k \in \mathcal{K},
\end{IEEEeqnarray}
whose number of delay spread taps is given by $L_k = \max(L^d_k, L^r_k)$. We assume that the number of cyclic prefixes (CP) is no smaller than the maximum number of delay spread taps for all devices, so that the inter-symbol interference (ISI) can be eliminated. Upon denoting the $m$-th row of the $M \times M$ discrete Fourier transform (DFT) matrix $\pmb{F}_M$ by $\pmb{f}^H_m$, we formulate the channel frequency response (CFR) for the $k$-th device at the $m$-th sub-band as
\begin{IEEEeqnarray}{rCl}\label{eqn:composite_channel}
C_{k,m}(\pmb{\Theta}) = \pmb{f}^H_m \pmb{h}_k  =  \pmb{f}^H_m \pmb{h}^d_k + \pmb{f}^H_m \pmb{V}_k \pmb{\Theta}, \quad \forall k \in \mathcal{K}, \forall m \in \mathcal{M}.
\end{IEEEeqnarray}
For ease of exposition, we assume that the knowledge of $\pmb{h}^d_k$ and of $\pmb{V}_k$ is perfectly known at the HAP. Naturally, this assumption is idealistic. Hence, the algorithm developed in this paper can be deemed to represent the best-case bound for the energy performance of realistic scenarios.
Since different types of signals are transmitted in the WET and computing phases, the reflection coefficients of the IRS require separate designs in these two phases.
The models of the WET and of computing phases are detailed as follows.

\subsection{Model of the Wireless Energy Transfer Phase}
It is assumed that the capacity of devices' battery is large enough so that all the harvest energy can be saved without energy overflow.
Let us use $\pmb{\Theta}^E = \big\{ \Theta^E_1, \Theta^E_2, \ldots, \Theta^E_N \big\}$ to represent the IRS reflection-coefficient vector during the WET phase, where we have $|\Theta^E_n| \leq 1$ for $\forall n \in \mathcal{N}$. Then, the composite channel of the $m$-th sub-band for the $k$-th device during the WET phase $C_{k,m}(\pmb{\Theta}^E)$ can be obtained by \eqref{eqn:composite_channel}.
As a benefit of the broadcasting nature of WET, each device can harvest the energy from the RF signals transmitted over all $M$ sub-bands.
Hence,
upon denoting the power allocation for the energy-carrying RF signals at the $M$ sub-bands during the WET phase by $\pmb{p}^E = [p^E_{1}, p^E_{2}, \ldots , p^E_{M}]$,
we are readily to formulate the energy harvested by the $k$-th device as \cite{wang2018joint}
\begin{IEEEeqnarray}{rCl}\label{eqn:WET}
E_k(\tau, \pmb{p}^E, \pmb{\Theta}^E) = \sum^M_{m=1} \eta \tau T p^E_{m} \big|C_{k,m}(\pmb{\Theta}^E)\big|^2,
\end{IEEEeqnarray}
where $\eta \in [0,1]$ denotes the efficiency of the energy harvesting at the wireless devices.

\subsection{Model of the Computing Phase}
We consider the data-partitioning based application \cite{wang2016mobile}, where a fraction of the data can be processed locally, while the other part can be offloaded to the edge node.
For a specific time block, we use $L_k$ and $\ell_k$ to denote the number of bits to be processed by the $k$-th device and its computation offloading volume in terms of the number of bits, respectively. The models of local computing, of computation offloading and of edge computing are detailed as follows.
\subsubsection{Local Computing}
We use $f_k$ and $c_k$ to represent its computing capability in terms of the number of central processing unit (CPU) cycles per second and the number of CPU cycles required to process a single bit, for the $k$-th device, respectively.
The number of bits processed by local computing is readily calculated as $(1-\tau)Tf_k/c_k$, and the number of bits to be offloaded is given by $\ell_k = L_k - (1-\tau)Tf_k/c_k$.
Furthermore, we assume that $f_k$ is controlled in the range of $[0,f_{max}]$ using the dynamic voltage scaling model \cite{wang2016mobile}. Upon denoting the computation energy efficiency coefficient of the processor's chip by $\kappa$, we formulate the power consumption of the local computing mode as $\kappa f_k^2$ for the $k$-th device \cite{wang2016mobile}.

\subsubsection{Computation offloading}
In order to mitigate the co-channel interference, the devices' computation offloading is realized relying on the orthogonal frequency-division multiple access (OFDMA) scheme. In this case, each sub-band is allowed to be used by at most a single device.
We use  the binary vector $\pmb{\alpha}_k = [\alpha_{k,1},\alpha_{k,2},\ldots,\alpha_{k,M} ]^T$ and the non-negative vector $\pmb{p}^I_k = [p^I_{k,1},p^I_{k,2}, \ldots, p^I_{k,M}]^T$ to represent the association between the sub-band and devices as well as the power allocation of the $k$-th device to the $M$ sub-bands, respectively, where we have
\begin{align}
\alpha_{k,m}
& = \begin{cases}
0,  & \quad \text{if } p^I_{k,m} = 0, \\
1,  & \quad \text{if } p^I_{k,m} > 0.
\end{cases}
\end{align}
The power consumption of computation offloading is given by $\sum^M_{m=1} \alpha_{k,m}( p_{k,m} + p_c)$, where $p_c$ represents a constant circuit power (caused by the digital-to-analog converter, filter, etc.) \cite{wang2018joint}.
Let us denote the IRS reflection-coefficient vector during the computation offloading by $\pmb{\Theta}^I = \big\{\Theta^I_1, \Theta^I_2, \ldots, \Theta^I_N \big\}$, where $|\Theta^I_{n}| \leq 1$ for $\forall n\in \mathcal{N}$.
Then, the composite channel of the $k$-th device at the $m$-th sub-band denoted by $C_{k,m}(\pmb{\Theta}^I)$ can be obtained by \eqref{eqn:composite_channel}.
The corresponding achievable rate of computation offloading is formulated below for the $k$-th device
\begin{IEEEeqnarray}{rCl}\label{eqn:achievable_rate}
R_k(\pmb{\alpha}_k,\pmb{p}^I_k,\pmb{\Theta}^I)=\sum^M_{m=1} \alpha_{k,m} B \log_2 \bigg(1+\frac{p_{k,m}|C_{k,m}(\pmb{\Theta}^I)|^2}{\Gamma \sigma^2} \bigg),
\end{IEEEeqnarray}
where $\Gamma$ is the gap between the channel capacity and a specific modulation and coding scheme.
Furthermore, in order to offload all the $\ell_k$ bits within the duration of  the computation phase, the achievable offloading rate has to obey $R_k(\tau,\pmb{\alpha}_k,\pmb{p}^I_k,\pmb{\Theta}^I) \geq \frac{\ell_k}{(1-\tau)T}$.

\subsubsection{Edge Computing}
Invoking the simplified linear model \cite{wang2018joint}, we formulate the energy consumption at the edge node as
 $\vartheta \sum^K_{k=1} \ell_k = \vartheta \sum^K_{k=1} \big[L_k - (1-\tau)Tf_k/c_k\big]$.
Furthermore, the latency imposed by edge computing comprises of two parts. The first part is caused by processing the computational tasks. Given that edge nodes typically possess high computational capabilities, this part can be negligible. The second part is induced by sending back the computational results, which are usually of a small volume. Hence, the duration of sending the feedback is also negligible. As such, we neglect the latency induced by edge computing.

\subsection{Problem Formulation}
In this paper, we aim for minimizing the total energy consumption of the OFDM-based WP-MEC system, by optimizing the time allocation for WET and computing phases $\tau$, both the power allocation $\pmb{p}^E$ and the IRS reflection coefficients $\pmb{\Theta}^E$ at the WET phase, and the local CPU frequency at devices $\pmb{f}$, the sub-band-device association $\{\pmb{\alpha}_k\}$ and the power allocation $\{\pmb{p}_k\}$ as well as the IRS reflection coefficients $\pmb{\Theta}^I$ at the computing phase, subject to the energy constraint imposed by energy harvesting, the latency requirement of computation offloading  and the sub-band-device association constraint in OFDMA systems as well as the constraint on IRS reflection coefficients.
Since the batteries of all the wireless devices are replenished by the HAP, their energy consumption is covered by the energy consumption at the HAP during the WET phase.
Hence, the total energy consumption of the system is formulated as the summation of the energy consumption both of the WET at the HAP and of the edge computing, i.e. $\tau T \sum^M_{m=1} p^E_{m} + \vartheta \sum^K_{k=1} \big[L_k - (1-\tau)Tf_k/c_k\big]$. To this end, the energy minimization problem is readily formulated for our OFDM-based WP-MEC system as
\begin{subequations}
\begin{align}
& \mathcal{P}0:  \mathop {\min} \limits_{\substack{\tau,\pmb{p}^E,\pmb{\Theta}^E, \pmb{f},\\ \{\pmb{\alpha}_k \}, \{\pmb{p}^I_k\}, \pmb{\Theta}^I}} \tau T \sum^M_{m=1} p^E_{m} + \vartheta \sum^K_{k=1} \bigg[L_k - \frac{(1-\tau)Tf_k}{c_k}\bigg] \nonumber \\
& \text{s.t.} \quad 0 < \tau < 1,  \label{eqn:P1_constraint_3} \\
& \quad \quad p^E_{m} \geq 0, \quad \forall m\in \mathcal{M},
\label{eqn:P1_constraint_4} \\
& \quad \quad | \Theta^E_{n} | \leq 1, \quad \forall n \in \mathcal{N},
\label{eqn:P1_constraint_5} \\
& \quad \quad 0 \leq f_k \leq f_{max}, \quad \forall k \in \mathcal{K},
\label{eqn:P1_constraint_10} \\
& \quad \quad \alpha_{k,m} \in \{0,1 \}, \quad \forall k \in \mathcal{K}, \quad \forall m \in \mathcal{M},
\label{eqn:P1_constraint_6} \\
& \quad \quad \sum^K_{k=1} \alpha_{k,m} \leq 1, \quad \forall m \in \mathcal{M},
\label{eqn:P1_constraint_7}\\
& \quad \quad p^I_{k,m} \geq 0, \quad \forall k \in \mathcal{K}, \quad \forall m\in \mathcal{M},  \label{eqn:P1_constraint_8} \\
& \quad \quad |\Theta^I_{n} | \leq 1, \quad \forall n \in \mathcal{N},
\label{eqn:P1_constraint_9} \\
& \quad \quad (1-\tau)T \bigg[ \kappa f_k^2 + \sum^M_{m=1} \alpha_{k,m} (p^I_{k,m} + p_c) \bigg] \leq E_k(\tau, \pmb{p}^E, \pmb{\Theta}^E), \quad \forall k \in \mathcal{K},
\label{eqn:P1_constraint_1}\\
& \quad \quad (1-\tau)T R_k(\pmb{\alpha}_k,\pmb{p}^I_k,\pmb{\Theta}^I) \geq L_k - \frac{(1-\tau)Tf_k}{c_k}, \quad \forall k \in \mathcal{K}.
\label{eqn:P1_constraint_2}
\end{align}
\end{subequations}
Constraint \eqref{eqn:P1_constraint_3} restricts the time allocation for the WET and for the computing phases.
Constraint \eqref{eqn:P1_constraint_4} and \eqref{eqn:P1_constraint_5} represent the range of the power allocation and the IRS reflection coefficients at the WET phase, respectively.
Constraint \eqref{eqn:P1_constraint_10} gives the range of tunable local computing frequencies.
Constraint \eqref{eqn:P1_constraint_6} and \eqref{eqn:P1_constraint_7} detail the requirement of sub-band-device association in OFDMA systems.
Constraint \eqref{eqn:P1_constraint_8} and \eqref{eqn:P1_constraint_9} restrict the range of the power allocation and the IRS reflection coefficient at the computing phase, respectively.
Constraint \eqref{eqn:P1_constraint_1} indicates that the sum energy consumption of local computing and of computation offloading should not exceed the harvested energy for each device. Finally, Constraint \eqref{eqn:P1_constraint_2} implies that the communication link between the $k$-th device and the HAP is capable of offloading the corresponding computational tasks within the duration of the computing phase.

\section{Joint Optimization of the Settings in the WET and the Computing Phases}
\label{sec:solution}
In this section, we propose to solve Problem $\mathcal{P}0$ in a two-step procedure. Firstly, given a fixed $\hat{\tau} \in (0,1)$, Problem $\mathcal{P}0$ can be simplified as follows
\begin{subequations}
\begin{align}
& \mathcal{P}1:  \mathop {\min} \limits_{\substack{\pmb{p}^E,\pmb{\Theta}^E, \pmb{f},\\ \{\pmb{\alpha}_k \}, \{\pmb{p}^I_k\}, \pmb{\Theta}^I}} \hat{\tau} T \sum^M_{m=1} p^E_{m} + \vartheta \sum^K_{k=1} \bigg[L_k - \frac{(1-\hat{\tau})Tf_k}{c_k}\bigg] \nonumber \\
& \text{s.t.} \quad
\eqref{eqn:P1_constraint_4}, \eqref{eqn:P1_constraint_5}, \eqref{eqn:P1_constraint_10}, \eqref{eqn:P1_constraint_6},
\eqref{eqn:P1_constraint_7}, \eqref{eqn:P1_constraint_8}, \eqref{eqn:P1_constraint_9}, \eqref{eqn:P1_constraint_1},
\eqref{eqn:P1_constraint_2}
\end{align}
\end{subequations}
In the second step, we aim for finding the optimal $\hat{\tau}$ that is capable of minimizing the OF of Problem $\mathcal{P}0$ using the one-dimensional search method. In the rest of this section, we focus on solving Problem $\mathcal{P}1$.
At a glance of Problem $\mathcal{P}1$, the optimization variables $\pmb{f}$, $\{\pmb{\alpha}_k\}$ and $\{\pmb{p}^I_k\}$ are coupled with $\pmb{p}^E$ and $\pmb{\Theta}^E$ in Constraint \eqref{eqn:P1_constraint_1}, which makes the problem difficult to solve.
To tackle this issue, the AO technique is invoked.
Specifically, upon initializing the setting of the computing phase, we may optimize the design of the WET phase while fixing the time allocation and the computing phase settings. Then, the computing phase settings could be optimized while fixing the time allocation and the design of the WET. A suboptimal solution can be obtained by iteratively optimizing the designs of the WET and of the computing phases. Let us detail the initialization as well as the designs of the WET and of the computing phases, as follows.

\subsection{Initialization of the Time Allocation and the Computing Phase}
\label{sec:Initialization}
In order to ensure our WET design to be a feasible solution of Problem $\mathcal{P}1$, the initial settings of the computing phase denoted by $\pmb{f}^{(0)}, \big\{\pmb{\alpha}_k^{(0)} \big\}, \big\{{\pmb{p}^I_k}^{(0)}\big\}, {\pmb{\Theta}^I}^{(0)}$ should satisfy Constraint \eqref{eqn:P1_constraint_10}, \eqref{eqn:P1_constraint_6}, \eqref{eqn:P1_constraint_7}, \eqref{eqn:P1_constraint_8}, \eqref{eqn:P1_constraint_9} and \eqref{eqn:P1_constraint_2}. Without any loss of generality, their initialization is set as follows.
\begin{itemize}
\item Local computing frequency $\pmb{f}^{(0)}$: Obeying the uniform distribution, each element of $\pmb{f}^{(0)}$ is randomly set in the range of $[0,f_{max}]$.
\item IRS reflection coefficient at the computing phase ${\pmb{\Theta}^I}^{(0)}$: Obeying the uniform distribution, the amplitude response ${\beta^I_n}^{(0)}$ and the phase shift ${\theta^I_n}^{(0)}$ are randomly set in the range of $[0,1]$ and of $[0,2\pi)$, respectively. Then, ${\pmb{\Theta}^I}^{(0)} = \{{\beta^I_1}^{(0)} e^{j{\theta^I_1}^{(0)}}, \ldots, {\beta^I_N}^{(0)} e^{j{\theta^I_N}^{(0)}}\}$ can be obtained.
\item Sub-band-device association at the computing phase $\big\{\pmb{\alpha}_k^{(0)}\big\}$: We reserve a single sub-band for the devices associated with the index ranging from $k=1$ to $k = K$, sequentially. Specific to the $k$-th device, we use $k_m^{(0)}$ to denote the sub-band having the maximum $\big|C_{k,m}\big({\pmb{\Theta}^I}^{(0)}\big)\big|^2$ over the unassigned sub-bands, and assign this sub-band to the $k$-th device.
\item Power allocation at the computing phase $\big\{{\pmb{p}^I_k}^{(0)}\big\}$: For the $k$-th device, its power allocation at the computing phase should satisfy Constraint \eqref{eqn:P1_constraint_2}. For minimizing the energy consumption, we assume the equivalence of two sides in Constraint \eqref{eqn:P1_constraint_2}. Then, its initial power allocation is given by ${p^{I^{(0)}}_{k,k_m^{(0)}}} = \frac{\Gamma \sigma^2 \Big[  2^{\frac{L_k}{(1-\hat{\tau})TB} - \frac{f_k}{c_k B}} - 1 \Big]}{\big|c_{k,k^{(0)}_m}\big({\pmb{\Theta}^I}^{(0)}\big) \big|^2}$. For those sub-bands associated with the index $m \neq k^{(0)}_m$, we set ${p^{I^{(0)}}_{k,m}} = 0$.
\end{itemize}

\subsection{Design of the WET Phase While Fixing the Time Allocation and Computing Settings}

Given a fixed time allocation $\hat{\tau}$ and the settings of the computing phase $\pmb{f}$, $\{\pmb{\alpha}_k \}$, $\{\pmb{p}^I_k\}$ and $\pmb{\Theta}^I$, we may simplify Problem $\mathcal{P}1$ as
\begin{subequations}
\begin{align}
& \mathcal{P}2:  \mathop {\min} \limits_{\pmb{p}^E,\pmb{\Theta}^E} \hat{\tau} T \sum^M_{m=1} p^E_{m} \nonumber \\
& \text{s.t.} \quad \eqref{eqn:P1_constraint_4}, \eqref{eqn:P1_constraint_5}, \nonumber \\
& \quad \quad (1-\hat{\tau})T \bigg[ \kappa f_k^2 + \sum^M_{m=1} \alpha_{k,m} (p^I_{k,m} + p_c) \bigg]\leq \sum^M_{m=1} \eta \hat{\tau} T p^E_{m} \big|C_{k,m}(\pmb{\Theta}^E)\big|^2, \quad \forall k \in \mathcal{K}.
\label{eqn:P2_constraint_1}
\end{align}
\end{subequations}
Since Constraint \eqref{eqn:P2_constraint_1} is not jointly convex regarding $\pmb{p}^E$ and $\pmb{\Theta}^E$, we optimize one of these two variables while fixing the other in an iterative manner, relying on the AO technique, as follows.

\subsubsection{Optimizing the Power Allocation of the WET Phase While Fixing the Settings of the Time Allocation, the Computing Phase and the IRS Reflection Coefficient at the WET Phase}

Given an IRS phase shift design $\pmb{\Theta}^E$, Problem $\mathcal{P}2$ is simplified as
\begin{subequations}
\begin{align}
& \mathcal{P}2\text{-}1:  \mathop {\min} \limits_{\pmb{p}^E} \hat{\tau} T \sum^M_{m=1} p^E_{m} \nonumber \\
& \text{s.t.} \quad \eqref{eqn:P1_constraint_4}, \eqref{eqn:P2_constraint_1}.
\end{align}
\end{subequations}
It can be observed that Problem $\mathcal{P}2\text{-}1$ is a linear programming problem, which can be readily solved with the aid of the general implementation of interior-point methods, e.g. CVX \cite{cvx}. The complexity is given by $\sqrt{M+KM}M[M+KM^3+M(M+KM^2)+M^2]$ \cite{wang2014outage}, i.e. $\mathcal{O}(K^{1.5} M^{4.5})$.

\subsubsection{Optimizing the IRS Reflection Coefficient at the WET Phase While Fixing the Settings of the Time Allocation, the Computing Phase and the power Allocation at the WET Phase}\label{sec:IRS_reflection_design_WET}

Given a power allocation at the WET phase $\pmb{p}^E$, Problem $\mathcal{P}2$ becomes a feasibility-check problem, i.e.
\begin{subequations}
\begin{align}
& \mathcal{P}2\text{-}2:  \text{Find } \pmb{\Theta}^E  \nonumber \\
& \text{s.t.} \quad \eqref{eqn:P1_constraint_5}, \eqref{eqn:P2_constraint_1}.
\end{align}
\end{subequations}
As verified in \cite{wu2019intelligent}, if one of the sub-problems is a feasibility-check problem, the iterative algorithm relying on the AO technique has a slow convergence. Specific to the problem considered, the operation of Find in Problem $\mathcal{P}2\text{-}2$ cannot guarantee the OF of Problem $\mathcal{P}2$ to be further reduced in each iteration.
To address this issue, we reformulate Problem $\mathcal{P}2\text{-}2$ as follows, by introducing a set of auxiliary variables $\{\xi_k\}$
\begin{subequations}
\begin{align}
& \mathcal{P}2\text{-}2':  \mathop {\max} \limits_{\pmb{\Theta}^E,\{\xi_k\}} \sum^K_{k=1} \xi_k   \nonumber \\
& \text{s.t.} \quad \eqref{eqn:P1_constraint_5}, \nonumber \\
& \quad \quad \xi_k + \kappa f_k^2 + \sum^M_{m=1} \alpha_{k,m} (p^I_{k,m} + p_c) \leq  \frac{\sum^M_{m=1} \eta \hat{\tau} p^E_{m} \big|\pmb{f}^H_m \pmb{h}^d_k + \pmb{f}^H_m \pmb{V}_k \pmb{\Theta}^E\big|^2}{1-\hat{\tau}}, \quad \forall k \in \mathcal{K},
\label{eqn:P2_2_constraint_1} \\
& \quad \quad \xi_k \geq 0, \quad \forall k \in \mathcal{K}.
\label{eqn:P2_2_constraint_11}
\end{align}
\end{subequations}
It is readily seen that the energy harvested by the wireless devices may increase after solving Problem $\mathcal{P}2\text{-}2'$, which implies that the channel gain of the reflection link is enhanced. Then, a reduced power of energy signals can be guaranteed, when we turn back to solve Problem $\mathcal{P}2\text{-}1$. As such, a faster convergence can be obtained.
However, at a glance of Problem $\mathcal{P}2\text{-}2'$, Constraint \eqref{eqn:P2_2_constraint_1} is still non-convex regarding $\pmb{\Theta}^E$.
To tackle this issue, we manipulate the optimization problem in light of \cite{yang2019irs} as follows.
Firstly, we transform Problem $\mathcal{P}2\text{-}2'$ to its equivalent problem below, by introducing a set of auxiliary variables $\pmb{y}^E$, $\pmb{a}^E$ and $\pmb{b}^E$
\begin{subequations}
\begin{align}
& \mathcal{P}2\text{-}2'E1:  \mathop {\max} \limits_{\pmb{\Theta}^E,\{\xi_k\}, \pmb{y}^E, \pmb{a}^E, \pmb{b}^E} \sum^K_{k=1} \xi_k   \nonumber \\
& \text{s.t.} \quad \eqref{eqn:P1_constraint_5}, \eqref{eqn:P2_2_constraint_11}, \nonumber \\
& \quad \quad \xi_k + \kappa f_k^2 + \sum^M_{m=1} \alpha_{k,m} (p^I_{k,m} + p_c) \leq  \frac{\sum^M_{m=1} \eta \hat{\tau} p^E_{m} y^E_{k,m}}{1-\hat{\tau}}, \quad \forall k \in \mathcal{K},
\label{eqn:P2_2'E_constraint_1} \\
& \quad \quad a^E_{k,m} = \Re \big\{ \pmb{f}^H_m \pmb{h}^d_k + \pmb{f}^H_m \pmb{V}_k \pmb{\Theta}^E \big\}, \quad k \in \mathcal{K}, \quad m \in \mathcal{M},
\label{eqn:P2_2'E_constraint_6}\\
& \quad \quad b^E_{k,m} = \Im \big\{ \pmb{f}^H_m \pmb{h}^d_k + \pmb{f}^H_m \pmb{V}_k \pmb{\Theta}^E \big\}, \quad k \in \mathcal{K}, \quad m \in \mathcal{M},
\label{eqn:P2_2'E_constraint_7}\\
& \quad \quad y^E_{k,m} \leq (a^E_{k,m})^2 + (b^E_{k,m})^2, \quad k \in \mathcal{K}, \quad m \in \mathcal{M},
\label{eqn:P2_2'E_constraint_8}
\end{align}
\end{subequations}
where $\Re\{\bullet \}$ and $\Im\{\bullet \}$ represent the real and imaginary part of $\bullet$, respectively.
Following this, the successive convex approximation (SCA) method \cite{razaviyayn2014successive} is applied for tackling the non-convex constraint \eqref{eqn:P2_2'E_constraint_8}.
Specifically, the approximation function is constructed as follows. The right hand side of \eqref{eqn:P2_2'E_constraint_8} is lower-bounded by its first-order approximation at $(\tilde{a}^E_{k,m},\tilde{b}^E_{k,m})$, i.e. $(a^E_{k,m})^2 + (b^E_{k,m})^2 \geq \tilde{a}^E_{k,m} (2a^E_{k,m} - \tilde{a}^E_{k,m}) + \tilde{b}^E_{k,m}(2b^E_{k,m} - \tilde{b}^E_{k,m})$, where the equality holds only when we have $\tilde{a}^E_{k,m} = a^E_{k,m}$ and $\tilde{b}^E_{k,m} = b^E_{k,m}$.
Now we consider the following optimization problem
\begin{subequations}
\begin{align}
& \mathcal{P}2\text{-}2'E2:  \mathop {\max} \limits_{\pmb{\Theta}^E,\{\xi_k\}, \pmb{y}^E, \pmb{a}^E, \pmb{b}^E} \sum^K_{k=1} \xi_k   \nonumber \\
& \text{s.t.} \quad \eqref{eqn:P1_constraint_5}, \eqref{eqn:P2_2_constraint_11}, \eqref{eqn:P2_2'E_constraint_1}, \eqref{eqn:P2_2'E_constraint_6}, \eqref{eqn:P2_2'E_constraint_7}, \nonumber \\
& \quad \quad y^E_{k,m} = \tilde{a}^E_{k,m} (2a^E_{k,m} - \tilde{a}^E_{k,m}) + \tilde{b}^E_{k,m}(2b^E_{k,m} - \tilde{b}^E_{k,m}), \quad k \in \mathcal{K}, \quad m \in \mathcal{M}.
\label{eqn:P2_2'E2_constraint_8}
\end{align}
\end{subequations}
Both the OF and contraints in Problem $\mathcal{P}2\text{-}2'E2$ are affine. Hence, Problem $\mathcal{P}2\text{-}2'E2$ is a convex optimization problem, which can be solved by the implementation of interior-point methods, e.g. CVX \cite{cvx}. Then, a locally optimal solution of $\mathcal{P}2\text{-}2'$ can be approached by successively updating $\tilde{a}^E_{k,m}$ and $\tilde{b}^E_{k,m}$ based on the optimal solution of Problem $\mathcal{P}2\text{-}2'E2$, whose procedure is detailed in Algorithm~\ref{alg:SCA}. The computation complexity of the SCA method is analyzed as follows. Problem $\mathcal{P}2\text{-}2'E2$ involves $2KM$ linear equality constraints (equivalently $4KM$ inequality constraints) of size $2N+1$, $K$ linear inequality constraints of size $M+1$, $KM$ linear inequality constraints of size $3$, $K$ linear inequality constraints of size $1$, $N$ second-order cone inequality constraints of size $2$. Hence, the total complexity of Algorithm~\ref{alg:SCA} is given by $\ln(1/\epsilon)\sqrt{4KM(2N+1)+K(M+1)+3KM+K+2N}(2N+3M+K)\{4KM(2N+1)^3 + K(M+1)^3 + 27KM + K + (2N+3M+K) [4KM(2N+1)^2 + K(M+1)^2 + 9KM + K] + 4N + (2N+3M+K)^2 \}$ \cite{wang2014outage}, i.e. $\ln(1/\epsilon)\mathcal{O}(K^{1.5}M^{1.5}N^{4.5}+K^{1.5}M^{2.5}N^{3.5} + K^{1.5}M^{2.5}N^{1.5} + K^{2.5}M^{1.5}N^{3.5} + K^{1.5}M^{4.5} + K^{2.5}M^{2.5}N^{2.5} + K^{2.5}M^{3.5} + K^{3.5}M^{1.5}N^{2.5} + K^{3.5}M^{2.5})$. To this end, we summarize the procedure of solving Problem $\mathcal{P}2$ in Algorithm~\ref{alg:solution_Problem_P2}.

\begin{algorithm}[h]\small
\caption{SCA approach to design $\pmb{\Theta}^E$, given the settings of the time allocation, the computing phase and the power allocation at the WET phase}
\begin{algorithmic}
 \renewcommand{\algorithmicrequire}{\textbf{Input:}}
 \renewcommand{\algorithmicensure}{\textbf{Output:}}
 \REQUIRE $t_{max}$, $\epsilon$, $K$, $M$, $N$, $T$, $\eta$, $c_k$, $\kappa$, $f_{max}$, $p_c$, $\Gamma$, $L_k$, $\{\pmb{h}^d_k\}$, $\{\pmb{V}_k\}$, $\hat{\tau}$, $\pmb{P}^E$, $\pmb{f}$, $\{\pmb{\alpha}_k\}$, $\{\pmb{p}^I_k\}$, $\pmb{\Theta}^I$ and $\tilde{\pmb{\Theta}}^E$
 \ENSURE  $\pmb{\Theta}^E$
\\ \textbf{1. Initialization}
\STATE Initialize $t_1 = 0$; $\epsilon_1 = 1$; $\xi_k = 0, \forall k \in \mathcal{K}$
\\ \textbf{2. SCA approach to design $\pmb{\Theta}^E$}
\WHILE{$t_1 < t_{\text{max}}$ $\&\&$ $\epsilon_1 > \epsilon$}
\STATE $\bullet$ Set $\tilde{a}^E_{k,m} = \Re \big\{ \pmb{f}^H_m \pmb{h}^d_k + \pmb{f}^H_m \pmb{V}_k \tilde{\pmb{\Theta}}^E \big\}$ and $\tilde{b}^E_{k,m} = \Im \big\{ \pmb{f}^H_m \pmb{h}^d_k + \pmb{f}^H_m \pmb{V}_k \tilde{\pmb{\Theta}}^E \big\}, \forall k \in \mathcal{K}, \forall m \in \mathcal{M}$
\STATE $\bullet$ Obtain ${\pmb{\Theta}^E}$ and $\{ \xi_k \}$ by solving Problem $\mathcal{P}2\text{-}2'E2$ using CVX
\STATE $\bullet$ Set $\epsilon_1 =\frac{ \big|\text{obj}\big( {\pmb{\Theta}}^E \big)- \text{obj}\big( \tilde{\pmb{\Theta}}^E \big)\big|}{\big| \text{obj}\big( \pmb{\Theta}^E \big) \big|}$, $\tilde{\pmb{\Theta}}^E \leftarrow \pmb{\Theta}^E$, $t_1 \leftarrow t_1 + 1$
\ENDWHILE
\\ \textbf{3. Output optimal ${\pmb{\Theta}^E}^*$}
\STATE ${\pmb{\Theta}^E}^* \leftarrow \tilde{\pmb{\Theta}}^E$
\end{algorithmic}
\label{alg:SCA}
\end{algorithm}

\begin{algorithm}[h]\small
\caption{Alternative optimization of $\pmb{p}^E$ and $\pmb{\Theta}^E$, given the settings of the time allocation and the computing phase}
\begin{algorithmic}
 \renewcommand{\algorithmicrequire}{\textbf{Input:}}
 \renewcommand{\algorithmicensure}{\textbf{Output:}}
 \REQUIRE $t_{max}$, $\epsilon$, $K$, $M$, $N$, $T$, $\eta$, $c_k$, $\kappa$, $f_{max}$, $p_c$, $\Gamma$, $L_k$, $\{\pmb{h}^d_k\}$, $\{\pmb{V}_k\}$, $\hat{\tau}$, $\pmb{f}$, $\{\pmb{\alpha}_k\}$, $\{\pmb{p}^I_k\}$, $\pmb{\Theta}^I$ and $\tilde{\pmb{\Theta}}^E$
 \ENSURE $\pmb{P}^E$ and $\pmb{\Theta}^E$
\\ \textbf{1. Initialization}
\STATE $\bullet$ Initialize $t_2 = 0$; $\epsilon_2 = 1$; ${\pmb{\Theta}^E}^{(0)} = \tilde{\pmb{\Theta}}^E$
\STATE $\bullet$ Given ${\pmb{\Theta}^E}^{(0)}$, find ${\pmb{P}^E}^{(0)}$ by solving Problem $\mathcal{P}2\text{-}1$ via CVX
\\ \textbf{2. Alternative optimization of $\pmb{P}^E$ and $\pmb{\Theta}^E$}
\WHILE{$t_2 < t_{\text{max}}$ $\&\&$ $\epsilon_2 > \epsilon$}
\STATE $\bullet$ Given ${\pmb{P}^E}^{(t_2)}$ and $\tilde{\pmb{\Theta}}^E = {\pmb{\Theta}^E}^{(t_2)}$, find ${\pmb{\Theta}^E}^{(t_2+1)}$ by solving Problem $\mathcal{P}2\text{-}2'E1$ using Algorithm~\ref{alg:SCA}
\STATE $\bullet$ Given ${\pmb{\Theta}^E}^{(t_2+1)}$, find ${\pmb{P}^E}^{(t_2+1)}$ by solving Problem $\mathcal{P}2\text{-}1$ via CVX
\STATE $\bullet$ Set $\epsilon_2 =\frac{ \big|\text{obj}\big( {\pmb{p}^E}^{(t_2+1)}, {{\pmb{\Theta}}^E}^{(t_2+1)} \big)- \text{obj}\big( {\pmb{p}^E}^{(t_2)}, {{\pmb{\Theta}}^E}^{(t_2)} \big)\big|}{\big| \text{obj}\big( {\pmb{p}^E}^{(t_2+1)}, {{\pmb{\Theta}}^E}^{(t_2+1)} \big) \big|}$, $t_2 \leftarrow t_2 + 1$
\ENDWHILE
\\ \textbf{3. Output optimal ${\pmb{P}^E}^*$ and ${\pmb{\Theta}^E}^*$}
\STATE ${\pmb{\Theta}^E}^* \leftarrow {\pmb{\Theta}^E}^{(t_2)}$ and ${\pmb{P}^E}^* \leftarrow {\pmb{P}^E}^{(t_2)}$
\end{algorithmic}
\label{alg:solution_Problem_P2}
\end{algorithm}

\subsection{Design of the Computing Phase While Fixing the Time Allocation and WET Settings}
In this subsection, we aim for optimizing the design of the computing phase, while fixing the time allocation $\hat{\tau}$ and the WET settings $\pmb{p}^E$ and $\pmb{\Theta}^E$. In this case, we simplify Problem $\mathcal{P}1$ as
\begin{subequations}
\begin{align}
& \mathcal{P}3:  \mathop {\min} \limits_{\pmb{f}, \{\pmb{\alpha}_k \}, \{\pmb{p}^I_k\}, \pmb{\Theta}^I} \vartheta \sum^K_{k=1} \bigg[L_k - \frac{(1-\hat{\tau})Tf_k}{c_k}\bigg] \nonumber \\
& \text{s.t.} \quad \eqref{eqn:P1_constraint_10}, \eqref{eqn:P1_constraint_6}, \eqref{eqn:P1_constraint_7}, \eqref{eqn:P1_constraint_8}, \eqref{eqn:P1_constraint_9}, \eqref{eqn:P1_constraint_1}, \nonumber \\
& \quad \quad \sum^m_{m=1} \alpha_{k,m} B \log_2 \Bigg[1+\frac{p_{k,m}|C_{k,m} (\pmb{\Theta}^I)|^2}{\Gamma \sigma^2} \Bigg] \geq \frac{L_k - \frac{(1-\hat{\tau})Tf_k}{c_k}}{(1-\hat{\tau})T}, \quad \forall k \in \mathcal{K}.   \label{eqn:P3_constraint_2}
\end{align}
\end{subequations}
Constraint \eqref{eqn:P3_constraint_2} is not jointly convex regarding $\{\pmb{\alpha}_k\}$, $\{\pmb{p}^I_k\}$ and $\pmb{\Theta}^I$. Hence, it is difficult to find its globally optimal solution.
Alternatively, its suboptimal solution is provided by iteratively optimizing the $\pmb{f}$, $\{\pmb{\alpha}_k\}$, $\{\pmb{p}^I_k\}$ and $\pmb{\Theta}^I$, again relying on the AO technique, as follows.

\subsubsection{Alternative Optimization of the Sub-Band-Device Association and the Power Allocation as well as the IRS Reflection Coefficient at the Computing Phase}
Given a fixed local CPU frequency setting $\pmb{f}$, the OF of Problem $\mathcal{P}3$ becomes deterministic. In other words, the optimization of $\{\pmb{\alpha}_k\}$, $\{\pmb{p}^I_k\}$ and $\pmb{\Theta}^I$ seems not contributing to reducing the OF. However, this is not always true, because if a larger feasible set of $\pmb{f}$ can be obtained by optimizing $\{\pmb{\alpha}_k\}$, $\{\pmb{p}^I_k\}$ and $\pmb{\Theta}^I$, a reduced OF may be achieved when we turn back to optimize $\pmb{f}$. Based on this observation, we formulate the problem below, by introducing a set of auxiliary variables $\{\zeta_k \}$
\begin{subequations}
\begin{align}
& \mathcal{P}3\text{-}1:  \mathop {\max} \limits_{\{\zeta_k \}, \{\pmb{\alpha}_k \}, \{\pmb{p}^I_k\}, \pmb{\Theta}^I} \sum^K_{k=1} \zeta_k \nonumber \\
& \text{s.t.} \quad  \eqref{eqn:P1_constraint_6}, \eqref{eqn:P1_constraint_7}, \eqref{eqn:P1_constraint_8}, \eqref{eqn:P1_constraint_9}, \eqref{eqn:P3_constraint_2} \nonumber \\
& \quad \quad \zeta_k \geq 0, \quad \forall k \in \mathcal{K}, \label{eqn:P3_1_constraint_3} \\
& \quad \quad (1-\hat{\tau})T \bigg[ \kappa f_k^2 + \sum^M_{m=1} \alpha_{k,m} (p^I_{k,m} + p_c) + \zeta_k \bigg] \leq \sum^M_{m=1} \eta \hat{\tau} T p^E_{m} \big|C_{k,m}(\pmb{\Theta}^E)\big|^2, \quad \forall k \in \mathcal{K}. \label{eqn:P3_1_constraint_2}
\end{align}
\end{subequations}
As specified in \eqref{eqn:P3_1_constraint_3}, the auxiliary variables $\{\zeta_k \}$ are non-negative, and thus a non-smaller set of $\pmb{f}$ may be obtained after solving Problem $\mathcal{P}3\text{-}1$.
Given that Constraint \eqref{eqn:P3_constraint_2} is not jointly convex regarding $\{ \pmb{p}_k \}$ and $\pmb{\Theta}^I$, we optimize $\{\pmb{\alpha}_k\}$, $\{\pmb{p}^I_k\}$ and $\pmb{\Theta}^I$ in two steps iteratively.

In the first step, we optimize $\{\zeta_k \}, \{\pmb{\alpha}_k \}$ and $\{\pmb{p}^I_k\}$, while fixing the IRS reflection coefficient $\pmb{\Theta}^I$. In this case, Problem $\mathcal{P}3\text{-}1$ can be simplified as
\begin{subequations}
\begin{align}
& \mathcal{P}3\text{-}1a:  \mathop {\max} \limits_{\{\zeta_k \}, \{\pmb{\alpha}_k \}, \{\pmb{p}^I_k\} } \sum^K_{k=1} \zeta_k \nonumber \\
& \text{s.t.} \quad  \eqref{eqn:P1_constraint_6}, \eqref{eqn:P1_constraint_7}, \eqref{eqn:P1_constraint_8},\eqref{eqn:P3_constraint_2}, \eqref{eqn:P3_1_constraint_3}, \eqref{eqn:P3_1_constraint_2}.
\end{align}
\end{subequations}
Problem $\mathcal{P}3\text{-}1a$ is a combinatorial optimization problem, where the binary constraint \eqref{eqn:P1_constraint_6} is non-convex.
The classic solution typically relies on the convex relaxation method \cite{wong1999multiuser}, where the binary constraint imposed on $\{\pmb{\alpha}_k\}$ is relaxed into a convex constraint by introducing time-sharing variables.
However, the relaxed problem is different from the original problem, which might lead to a specific error.
To address this issue, a near-optimal solution based on the Lagrangian duality was proposed \cite{seong2006optimal}, where it is verified that the duality gap vanishes in the system equipped with more than $8$ sub-bands. Hence, in this paper, the Lagrangian duality method \cite{boyd2004convex} is invoked for solving Problem $\mathcal{P}3\text{-}1a$.
Specifically, denoting the non-negative Lagrange multiplier vectors by $\pmb{\lambda} = [\lambda_1, \lambda_2, \ldots, \lambda_K]^T$ and $\pmb{\mu} = [ \mu_1, \mu_2, \ldots, \mu_K]^T$, we formulate the Lagrangian function of Problem $\mathcal{P}3\text{-}1a$ over the domain $\mathcal{D}$ as
\begin{IEEEeqnarray}{rCl}\label{eqn:lagrangian}
\mathcal{L} \big(\{\zeta_k\}, \{\pmb{p}^I_k\}, \pmb{\lambda}, \pmb{\mu} \big)
& = & \sum_{k=1}^K \zeta_k - \sum_{k=1}^K \lambda_k \bigg[ \kappa f_k^2 + \sum^M_{m=1} (p^I_{k,m} + p_c) + \zeta_k - \frac{E_k(\hat{\tau}, \pmb{p}^E, \pmb{\Theta}^E)}{(1-\hat{\tau})T} \bigg] \nonumber \\
& & + \sum_{k=1}^K \mu_k \Bigg[ \sum^M_{m=1} B \log_2 \bigg(1+\frac{p^I_{k,m}|C_{k,m} (\pmb{\Theta}^I)|^2}{\Gamma \sigma^2} \bigg) - \frac{L_k - \frac{(1-\hat{\tau})Tf_k}{c_k}}{(1-\hat{\tau})T} \Bigg],
\end{IEEEeqnarray}
where the domain $\mathcal{D}$ is defined as the set of all non-negative $p^I_{k,m}$ for $\forall k \in \mathcal{K}$ and for $\forall m \in \mathcal{M}$ such that for each $m$, only a single $p^I_{k,m}$ is positive for $k \in \mathcal{K}$. Then, the Lagrangian dual function of Problem $\mathcal{P}3\text{-}1a$ is given by
\begin{IEEEeqnarray}{rCl}\label{eqn:dual}
g(\pmb{\lambda},\pmb{\mu}) = \mathop {\max} \limits_{\{ \zeta_k \}, \{\pmb{p}^I_k \} \in {\mathcal{D}} } \mathcal{L} \big(\{\zeta_k\}, \{\pmb{p}^I_k\}, \pmb{\lambda}, \pmb{\mu} \big).
\end{IEEEeqnarray}
\eqref{eqn:dual} can be reformulated as
\begin{IEEEeqnarray}{rCl}\label{eqn:dual_2}
g(\pmb{\lambda},\pmb{\mu}) & = &  \sum_{m=1}^M \hat{g}_m(\pmb{\lambda},\pmb{\mu}) + \sum^K_{k=1} (1-\lambda_k) \zeta_k  -\sum^K_{k=1} \lambda_k \kappa f_k^2  \nonumber \\
& & + \sum^K_{k=1} \lambda_k \frac{E_k(\hat{\tau}, \pmb{p}^E, \pmb{\Theta}^E)}{(1-\hat{\tau})T} - \sum^K_{k=1}  \frac{ \mu_k\Big[ L_k - \frac{(1-\hat{\tau})Tf_k}{c_k} \Big]}{(1-\hat{\tau})T},
\end{IEEEeqnarray}
where we have
\begin{IEEEeqnarray}{rCl}\label{eqn:hat_g_m}
\hat{g}_m(\pmb{\lambda},\pmb{\mu}) \triangleq \mathop {\max} \limits_{\{\pmb{p}^I_k \} \in {\mathcal{D}} }
\Bigg\{ - \sum_{k=1}^K \lambda_k (p^I_{k,m} + p_c) + \sum_{k=1}^K \mu_k B \log_2 \Bigg[1+\frac{p^I_{k,m}|C_{k,m} (\pmb{\Theta}^I)|^2}{\Gamma \sigma^2} \Bigg] \Bigg\}.
\end{IEEEeqnarray}
It is readily seen that \eqref{eqn:hat_g_m} is concave regarding $p^I_{k,m}$. Thus, upon letting its first-order derivative regarding $p^I_{k,m}$ be $0$, we may give the optimal power of the $m$-th sub-band when it is allocated to the $k$-th device as
\begin{IEEEeqnarray}{rCl}\label{eqn:p_k_m}
\hat{p}^{I}_{k,m}(\lambda_k,\mu_k) = \bigg[ \frac{\mu_k B}{\lambda_k \ln 2} - \frac{\Gamma \sigma^2}{|C_{k,m} (\pmb{\Theta}^I)|^2} \bigg]^+.
\end{IEEEeqnarray}
Then, $\hat{g}_m(\pmb{\lambda},\pmb{\mu})$ can be obtained, by searching over all possible assignments of the $m$-th sub-band for all the $K$ devices, as follows
\begin{IEEEeqnarray}{rCl}\label{eqn:alpha_k_m}
\hat{g}_m(\pmb{\lambda},\pmb{\mu}) = \max_k \Bigg\{ - \lambda_k \Big[ \hat{p}^{I}_{k,m}(\lambda_k,\mu_k) + p_c \Big] + \mu_k B \log_2 \Bigg[1+\frac{\hat{p}^{I}_{k,m}(\lambda_k,\mu_k)|C_{k,m} (\pmb{\Theta}^I)|^2}{\Gamma \sigma^2} \Bigg] \Bigg\},
\end{IEEEeqnarray}
and the suitable device is given by $k^* = \arg \hat{g}_m(\pmb{\lambda},\pmb{\mu})$. We set $\alpha_{k^*,m} =1$ and $p^{I}_{k^*,m} = \hat{p}^I_{k^*,m}$ as well as $\alpha_{k,m} =0$ and $p^{I}_{k,m} = 0$ for $\forall k \neq k^*$.
We continue by calculating $\{ \zeta_k \}$ as follows.
At a glance of \eqref{eqn:p_k_m}, it is observed that $\lambda_k$ has to yield $\lambda_k > 0$, $\forall k \in \mathcal{K}$, which implies that Constraint \eqref{eqn:P3_1_constraint_2} is strictly binding for the optimal solution of Problem $\mathcal{P}3\text{-}1a$. Therefore, $\zeta_k$ can be set as
\begin{IEEEeqnarray}{rCl}\label{eqn:zeta_k}
\zeta_k = \frac{E_k(\hat{\tau}, \pmb{p}^E, \pmb{\Theta}^E)}{(1-\hat{\tau})T} - \kappa f_k^2 - \sum^M_{m=1} \alpha_{k,m} (p^I_{k,m} + p_c).
\end{IEEEeqnarray}
Once all $\hat{g}_m(\pmb{\lambda},\pmb{\mu})$ and $\zeta_k$ are obtained, $g(\pmb{\lambda},\pmb{\mu})$ can be calculated by \eqref{eqn:dual_2}.
Bearing in mind that the obtained $g(\pmb{\lambda},\pmb{\mu})$ is not guaranteed to be optimal, we have to find a suitable set of $\pmb{\lambda}$ and $\pmb{\mu}$ that minimize $g(\pmb{\lambda},\pmb{\mu})$, which can be realized by the ellipsoid method \cite{boyd2004convex}. More explicitly, the Lagrange multipliers are iteratively updated following their sub-gradients towards their optimal settings. The corresponding sub-gradients are given as follows
\begin{align}
& s_{\lambda_k} = \kappa f_k^2 + \sum^M_{m=1} \alpha_{k,m} (p^I_{k,m} + p_c) - \frac{E_k(\hat{\tau}, \pmb{p}^E, \pmb{\Theta}^E)}{(1-\hat{\tau})T}, \\
& s_{\mu_k} = \frac{L_k - \frac{(1-\hat{\tau})Tf_k}{c_k}}{(1-\hat{\tau})T} - \sum_{m=1}^M \alpha_{k,m} B \log_2 \bigg(1+\frac{p^I_{k,m}|C_{k,m} (\pmb{\Theta}^I)|^2}{\Gamma \sigma^2} \bigg).
\end{align}
Upon denoting the iteration index by $t$, the Lagrange multipliers are updated obeying $\lambda_k(t+1) = [\lambda_k(t) + \delta_{\lambda}(t) s_{\lambda_k}]^+$ and $\mu_k(t+1) = [\mu_k(t) + \delta_{\mu}(t) s_{\mu_k}]^+$, where we set $\delta_{\lambda}(t) = \delta_{\lambda}(1)/t$ and $\delta_{\mu}(t) = \delta_{\mu}(1)/t$ for ensuring the convergence of the OF. In the problem considered, the ellipsoid method converges in $\mathcal{O}(K^2)$ iterations \cite{boyd2004convex,seong2006optimal}. Within each iteration, the computational complexity is of $\mathcal{O}(KM)$. Hence, the total computational complexity is given by $\mathcal{O}(MK^3)$.
The procedure of this Lagrangian duality method is summarized in Algorithm~\ref{alg:dual_decomposition}.

\begin{algorithm}[h]\small
\caption{Design of $\{ \pmb{\alpha}_k \}$ and $\{ \pmb{p}^I_k \}$, given the settings of $\hat{\tau}$, $\pmb{p}^E$, $\pmb{\Theta}^E$, $\pmb{f}$ and $\pmb{\Theta}^I$}
\begin{algorithmic}
 \renewcommand{\algorithmicrequire}{\textbf{Input:}}
 \renewcommand{\algorithmicensure}{\textbf{Output:}}
 \REQUIRE $t_{max}$, $\epsilon$, $K$, $M$, $N$, $T$, $\eta$, $c_k$, $\kappa$, $f_{max}$, $p_c$, $\Gamma$, $L_k$, $\{\pmb{h}^d_k\}$, $\{\pmb{V}_k\}$, $\hat{\tau}$, $\pmb{P}^E$, $\pmb{\Theta}^E$, $\pmb{\Theta}^I$, $\pmb{f}$, $\pmb{\Theta}^I$, $\pmb{\lambda}$ and $\pmb{\mu}$
 \ENSURE $\{\zeta_k \}$, $\{\pmb{\alpha}_k\}$, $\{\pmb{p}^I_k\}$
\\ \textbf{1. Initialization}
\STATE Initialize $t_3 = 0$; $\epsilon_3 = 1$; Calculate $\mathcal{L}^{(0)}$ using \eqref{eqn:lagrangian}
\\ \textbf{2. Optimization of $\{\zeta_k \}$, $\{ \pmb{\alpha}_k \}$ and $\{ \pmb{p}^I_k \}$}
\WHILE{$t_3 < t_{\text{max}}$ $\&\&$ $\epsilon_3 > \epsilon$}
\FOR{$m=1:M$}
\STATE $\bullet$ Calculate $\hat{p}^{I}_{k,m}$ using \eqref{eqn:p_k_m} for each $\forall k \in \mathcal{K}$
\STATE $\bullet$ Obtain the optimal device $k^* = \arg \hat{g}_m(\pmb{\lambda},\pmb{\mu})$ in \eqref{eqn:alpha_k_m}
\STATE $\bullet$ Set $\alpha_{k^*,m} =1$ and $p^{I}_{k^*,m} = \hat{p}^I_{k^*,m}$ as well as $\alpha_{k,m} =0$ and $p^{I}_{k,m} = 0$ for $\forall k \neq k^*$
\ENDFOR
\STATE $\bullet$ Calculate $\zeta_k$ using \eqref{eqn:zeta_k}
\STATE $\bullet$ Calculate $\mathcal{L}^{(t_3+1)}$ using \eqref{eqn:lagrangian}
\STATE $\bullet$ Update $\pmb{\lambda}$ and $\pmb{\mu}$ using the ellipsoid method
\STATE $\bullet$ Set $\epsilon_3 =\frac{ \big|\mathcal{L}^{(t_3+1)}- \mathcal{L}^{(t_3)}\big|}{\big| \mathcal{L}^{(t_3+1)} \big|}$, $t_3 \leftarrow t_3 + 1$
\ENDWHILE
\\ \textbf{3. Output optimal $\{ \zeta_k \}^*$, $\{\pmb{\alpha}_k\}^*$ and $\{\pmb{p}^I_k\}^*$}
\STATE $\{ \zeta_k \}^* = \{ \zeta_k \} $, $\{\pmb{\alpha}_k\}^* = \{\pmb{\alpha}_k\}$, $\{\pmb{p}^I_k\}^* = \{\pmb{p}^I_k\}$
\end{algorithmic}
\label{alg:dual_decomposition}
\end{algorithm}

In the second step, we optimize the IRS reflection coefficient $\pmb{\Theta}^I$, while fixing the settings of the resource allocation at the computing phase $\{\pmb{\alpha}_k\}$ and $\{ \pmb{p}^I \}$. In this case, Problem $\mathcal{P}3\text{-}1$ becomes a feasibility-check problem below
\begin{subequations}
\begin{align}
& \mathcal{P}3\text{-}1b:  \text{Find } \pmb{\Theta}^I  \nonumber \\
& \text{s.t.} \quad \eqref{eqn:P1_constraint_9}, \eqref{eqn:P3_constraint_2}. \nonumber
\label{eqn:P3_2_constraint_9}
\end{align}
\end{subequations}
The problem can be solved using the approach devised in Section~\ref{sec:IRS_reflection_design_WET}, detailed as follows. By introducing a set of auxiliary variables $\{ \chi_k \}$, we transform $\mathcal{P}3\text{-}2$ to the problem below
\begin{subequations}
\begin{align}
& \mathcal{P}3\text{-}1b:  \mathop {\max} \limits_{\pmb{\Theta}^I,\{\chi_k\}} \sum^K_{k=1} \chi_k    \nonumber \\
& \text{s.t.} \quad \eqref{eqn:P1_constraint_9}, \nonumber \\
& \quad \quad \chi_k \geq 0, \quad \forall k \in \mathcal{K},
\label{eqn:P3_2'_constraint_3} \\
& \quad \quad \sum^m_{m=1} \alpha_{k,m} B \log_2 \bigg[1+\frac{p_{k,m}|C_{k,m} (\pmb{\Theta}^I)|^2}{\Gamma \sigma^2} \bigg] \geq \frac{L_k - \frac{(1-\hat{\tau})Tf_k}{c_k}}{(1-\hat{\tau})T} + \chi_k, \quad \forall k \in \mathcal{K}.   \label{eqn:P3_2'_constraint_2}
\end{align}
\end{subequations}
Constraint \eqref{eqn:P3_2'_constraint_2} is non-convex regarding $\pmb{\Theta}^I$. To address this issue, firstly we transform Problem $\mathcal{P}3\text{-}1b$ to its equivalent form, by introducing a set of auxiliary variables $\pmb{y}^I$, $\pmb{a}^I$ and $\pmb{b}^I$
\begin{subequations}
\begin{align}
& \mathcal{P}3\text{-}1bE1:  \mathop {\max} \limits_{\pmb{\Theta}^I,\{\chi_k\},\pmb{y}^I,\pmb{a}^I,\pmb{b}^I} \sum^K_{k=1} \chi_k    \nonumber \\
& \text{s.t.}  \quad \eqref{eqn:P1_constraint_9}, \eqref{eqn:P3_2'_constraint_3}, \nonumber \\
& \quad \quad \sum^m_{m=1} \alpha_{k,m} B \log_2 \bigg(1+\frac{p_{k,m}y^I_{k,m}}{\Gamma \sigma^2} \bigg) \geq \frac{L_k - \frac{(1-\hat{\tau})Tf_k}{c_k}}{(1-\hat{\tau})T} + \chi_k, \quad \forall k \in \mathcal{K},   \label{eqn:P3_2'E1_constraint_2} \\
& \quad \quad a^I_{k,m} = \Re \big\{ \pmb{f}^H_m \pmb{h}^d_k + \pmb{f}^H_m \pmb{V}_k \pmb{\Theta}^I \big\}, \quad k \in \mathcal{K}, \quad m \in \mathcal{M},
\label{eqn:P3_2'E1_constraint_6}\\
& \quad \quad b^I_{k,m} = \Im \big\{ \pmb{f}^H_m \pmb{h}^d_k + \pmb{f}^H_m \pmb{V}_k \pmb{\Theta}^I \big\}, \quad k \in \mathcal{K}, \quad m \in \mathcal{M},
\label{eqn:P3_2'E1_constraint_7}\\
& \quad \quad y^I_{k,m} = (a^I_{k,m})^2 + (b^I_{k,m})^2, \quad k \in \mathcal{K}, \quad m \in \mathcal{M}.
\label{eqn:P3_2'E1_constraint_8}
\end{align}
\end{subequations}
Then, upon invoking the so-called SCA method as detailed in Section~\ref{sec:IRS_reflection_design_WET}, we approach the locally optimal solution by solving the problem below in a successive manner
\begin{subequations}
\begin{align}
& \mathcal{P}3\text{-}1bE2:  \mathop {\max} \limits_{\pmb{\Theta}^I,\{\chi_k\}} \sum^K_{k=1} \chi_k    \nonumber \\
& \text{s.t.} \quad \eqref{eqn:P1_constraint_9}, \eqref{eqn:P3_2'_constraint_3}, \eqref{eqn:P3_2'E1_constraint_2}, \eqref{eqn:P3_2'E1_constraint_6}, \eqref{eqn:P3_2'E1_constraint_7}, \nonumber \\
& \quad \quad y^I_{k,m} = \tilde{a}^I_{k,m} (2a^I_{k,m} - \tilde{a}^I_{k,m}) + \tilde{b}^I_{k,m}(2b^I_{k,m} - \tilde{b}^I_{k,m}), \quad k \in \mathcal{K}, \quad m \in \mathcal{M}.
\label{eqn:P3_2'E2_constraint_8}
\end{align}
\end{subequations}
Problem $\mathcal{P}3\text{-}1bE2$ is a convex optimization problem, which can be readily solved with the aid of the software of CVX \cite{cvx}. The computational complexity is the same as that given in Section~\ref{sec:IRS_reflection_design_WET}. Note that the optimization of $\{ \pmb{\alpha}_k \}$, $\{ \pmb{p}^I_k \}$ and $\pmb{\Theta}^I$ not only contributes to reducing the OF of Problem $\mathcal{P}2$, but also leads to a decreased OF of Problem $\mathcal{P}1$ by slacking its constraint \eqref{eqn:P2_constraint_1}. Hence, we may still reduce the OF of Problem $\mathcal{P}1$ by iteratively optimizing the settings of the WET phase and the computing phase, even if $\pmb{f}$ reaches its maximum value.

\subsubsection{Design of CPU Frequencies}
Given the settings of the sub-band-device association $\{ \pmb{\alpha}_k \}$, the power allocation $\{\pmb{p}^I_k \}$ and the IRS reflection coefficient $\pmb{\Theta}^I$, Problem $\mathcal{P}3$ can be simplified as
\begin{subequations}
\begin{align}
& \mathcal{P}3\text{-}2:  \mathop {\min} \limits_{\pmb{f}, \{\pmb{\alpha}_k \}, \{\pmb{p}^I_k\}, \pmb{\Theta}^I} \vartheta \sum^K_{k=1} \bigg[L_k - \frac{(1-\hat{\tau})Tf_k}{c_k}\bigg] \nonumber \\
& \text{s.t.} \quad \eqref{eqn:P1_constraint_10}, \eqref{eqn:P3_1_constraint_2}.
\end{align}
\end{subequations}
It can be seen that the OF of Problem $\mathcal{P}3\text{-}2$ decreases upon increasing $\pmb{f}$. Hence, upon denoting
\begin{IEEEeqnarray}{rCl}
\hat{f}_k = \sqrt{\frac{\frac{E_k(\hat{\tau}, \pmb{p}^E, \pmb{\Theta}^E)}{(1-\hat{\tau})T} - \sum^M_{m=1} \alpha_{k,m} (p^I_{k,m} + p_c) - \zeta_k}{\kappa}},
\end{IEEEeqnarray}
the optimal $\pmb{f}$ can be obtained as:
\begin{align}\label{eqn:f_k}
f_k
& = \begin{cases}
0,  & \quad \text{if } \frac{E_k(\hat{\tau}, \pmb{p}^E, \pmb{\Theta}^E)}{(1-\hat{\tau})T} - \sum^M_{m=1} \alpha_{k,m} (p^I_{k,m} + p_c) - \zeta_k < 0, \\
\hat{f}_k, & \quad \text{if } 0 \leq \frac{E_k(\hat{\tau}, \pmb{p}^E, \pmb{\Theta}^E)}{(1-\hat{\tau})T} - \sum^M_{m=1} \alpha_{k,m} (p^I_{k,m} + p_c) - \zeta_k < \kappa f_{max}^2, \\
f_{max},  & \quad \text{if } \frac{E_k(\hat{\tau}, \pmb{p}^E, \pmb{\Theta}^E)}{(1-\hat{\tau})T} - \sum^M_{m=1} \alpha_{k,m} (p^I_{k,m} + p_c) - \zeta_k \geq \kappa f_{max}^2.
\end{cases}
\end{align}
The procedure of optimizing $\{ \pmb{\alpha}_k \}$, $\{ \pmb{p}^I_k \}$, $\pmb{\Theta}^I$ and $\pmb{f}$ is summarized in Algorithm~\ref{alg:solution_Problem_P3}. To this end, it is readily to summarize the algorithm solving Problem~$\mathcal{P}1$ under a given $\hat{\tau}$ in Algorithm~\ref{alg:solution_Problem_P}, and an appropriate $\tau$ is found with the aid of numerical results, as detailed in Section~\ref{sec:Selection of the Time Allocation}.

\begin{algorithm}[h]\small
\caption{Alternative optimization of $\pmb{f}$, $\{ \pmb{\alpha}_k \}$, $\{ \pmb{p}^I_k \}$ and $\pmb{\Theta}^I$, given the settings of $\hat{\tau}$, $\pmb{p}^E$ and $\pmb{\Theta}^E$}
\begin{algorithmic}
 \renewcommand{\algorithmicrequire}{\textbf{Input:}}
 \renewcommand{\algorithmicensure}{\textbf{Output:}}
 \REQUIRE $t_{max}$, $\epsilon$, $K$, $M$, $N$, $T$, $\eta$, $c_k$, $\kappa$, $f_{max}$, $p_c$, $\Gamma$, $L_k$, $\{\pmb{h}^d_k\}$, $\{\pmb{V}_k\}$, $\hat{\tau}$, $\pmb{P}^E$, $\pmb{\Theta}^E$, $\pmb{f}$, and $\tilde{\pmb{\Theta}}^I$
 \ENSURE $\{ \pmb{\alpha}_k \}$ $\{ \pmb{p}^I_k \}$ and $\pmb{\Theta}^I$
\\ \textbf{1. Initialization}
\STATE $\bullet$ Initialize $t_4 = 0$; $\epsilon_4 = 1$; ${\pmb{\Theta}^I}^{(0)} = \tilde{\pmb{\Theta}}^I$
\STATE $\bullet$ Given ${\pmb{\Theta}^I}^{(0)}$, find $\{ \pmb{\alpha}_k \}^{(0)}$ and $\{ \pmb{p}^I_k \}^{(0)}$ by solving Problem $\mathcal{P}3\text{-}1a$ via Algorithm~\ref{alg:dual_decomposition}
\STATE $\bullet$ Obtain $\pmb{f}^{(0)}$ via \eqref{eqn:f_k} and calculate $\text{obj}\big( \pmb{f}^{(0)} \big)$
\\ \textbf{2. Alternative optimization of $\pmb{f}$, $\{ \pmb{\alpha}_k \}$, $\{ \pmb{p}^I_k \}$ and $\pmb{\Theta}^I$}
\WHILE{$t_4 < t_{\text{max}}$ $\&\&$ $\epsilon_4 > \epsilon$}
\STATE $\bullet$ Given $\{ \pmb{\alpha}_k \}^{(t_4)}$, $\{ \pmb{p}^I_k \}^{(t_4)}$ and $\tilde{\pmb{\Theta}}^I = {\pmb{\Theta}^I}^{(t_4)}$, find ${\pmb{\Theta}^I}^{(t_4+1)}$ by solving Problem $\mathcal{P}3\text{-}1bE1$ via Algorithm~\ref{alg:SCA}
\STATE $\bullet$ Given ${\pmb{\Theta}^I}^{(t_4+1)}$, find $\{ \pmb{\alpha}_k \}^{(t_4+1)}$ and $\{ \pmb{p}^I_k \}^{(t_4+1)}$ by solving Problem $\mathcal{P}3\text{-}1a$ via Algorithm~\ref{alg:dual_decomposition}
\STATE $\bullet$ Obtain $\pmb{f}^{(t_4+1)}$ via \eqref{eqn:f_k} and calculate $\text{obj}\big( \pmb{f}^{(t_4+1)} \big)$
\STATE $\bullet$ Set $\epsilon_4 =\frac{ \big|\text{obj}\big( \pmb{f}^{(t_4+1)} \big)- \text{obj}\big( \pmb{f}^{(t_4)} \big)\big|}{\big| \text{obj}\big( \pmb{f}^{(t_4+1)} \big) \big|}$, $t_4 \leftarrow t_4 + 1$
\ENDWHILE
\\ \textbf{3. Output optimal $\{ \pmb{\alpha}_k \}^*$ $\{ \pmb{p}^I_k \}^*$ and ${\pmb{\Theta}^I}^*$}
\STATE $\{ \pmb{\alpha}_k \}^* \leftarrow \{ \pmb{\alpha}_k \}^{(t_4)}$, $\{ \pmb{p}^I_k \}^* \leftarrow \{ \pmb{p}^I_k \}^{(t_4)}$ and ${\pmb{\Theta}^I}^* \leftarrow {\pmb{\Theta}^I}^{(t_4)}$
\end{algorithmic}
\label{alg:solution_Problem_P3}
\end{algorithm}

\begin{algorithm}[h]\small
\caption{Alternative optimization of the WET and computing phases, given the time allocation}
\begin{algorithmic}
 \renewcommand{\algorithmicrequire}{\textbf{Input:}}
 \renewcommand{\algorithmicensure}{\textbf{Output:}}
 \REQUIRE $t_{max}$, $\epsilon$, $K$, $M$, $N$, $T$, $\eta$, $c_k$, $\kappa$, $f_{max}$, $p_c$, $\Gamma$, $L_k$, $\{\pmb{h}^d_k\}$, $\{\pmb{V}_k\}$ and $\hat{\tau}$
 \ENSURE $\pmb{P}^E$, $\pmb{\Theta}^E$, $\pmb{f}$, $\{ \pmb{\alpha}_k \}$ $\{ \pmb{p}^I_k \}$ and $\pmb{\Theta}^I$
\\ \textbf{1. Initialization}
\STATE $\bullet$ Initialize $t_5 = 0$; $\epsilon_5 = 1$; $\tilde{\pmb{\Theta}}^E$
\STATE $\bullet$ Initialize $\pmb{f}^{(0)}$, $\{ \pmb{\alpha}_k \}^{(0)}$, $\{ \pmb{p}^I_k \}^{(0)}$ and ${\pmb{\Theta}^I}^{(0)}$ following Section~\ref{sec:Initialization}
\STATE $\bullet$ Given $\pmb{f}^{(0)}$, $\{ \pmb{\alpha}_k \}^{(0)}$, $\{ \pmb{p}^I_k \}^{(0)}$ and ${\pmb{\Theta}^I}^{(0)}$, find ${\pmb{P}^E}^{(0)}$ and ${\pmb{\Theta}^E}^{(0)}$ by solving Problem $\mathcal{P}2$ via Algorithm~\ref{alg:solution_Problem_P2}
\\ \textbf{2. Alternative optimization of $\pmb{P}^E$, $\pmb{\Theta}^E$, $\pmb{f}$, $\{ \pmb{\alpha}_k \}$ $\{ \pmb{p}^I_k \}$ and $\pmb{\Theta}^I$}
\WHILE{$t_5 < t_{\text{max}}$ $\&\&$ $\epsilon_5 > \epsilon$}
\STATE $\bullet$ Given ${\pmb{P}^E}^{(t_5)}$, ${\pmb{\Theta}^E}^{(t_5)}$ and $\tilde{\pmb{\Theta}}^I = {\pmb{\Theta}^I}^{(t_5)}$, find $\pmb{f}^{(t_5+1)}$, $\{ \pmb{\alpha}_k \}^{(t_5+1)}$, $\{ \pmb{p}^I_k \}^{(t_5+1)}$ and ${\pmb{\Theta}^I}^{(t_5+1)}$ by solving Problem $\mathcal{P}3$ using Algorithm~\ref{alg:solution_Problem_P3}
\STATE $\bullet$ Given $\pmb{f}^{(t_5+1)}$, $\{ \pmb{\alpha}_k \}^{(t_5+1)}$, $\{ \pmb{p}^I_k \}^{(t_5+1)}$, ${\pmb{\Theta}^I}^{(t_5+1)}$ and $\tilde{\pmb{\Theta}}^E = {\pmb{\Theta}^E}^{(t_5)}$, find ${\pmb{P}^E}^{(t_5+1)}$ and ${\pmb{\Theta}^E}^{(t_5+1)}$ by solving Problem $\mathcal{P}2$ via Algorithm~\ref{alg:solution_Problem_P2}
\STATE $\bullet$ Set $\epsilon_5 =\frac{ \big|\text{obj}^{(t_5+1)} - \text{obj}^{(t_5)}\big|}{\big| \text{obj}^{(t_5+1)} \big|}$, $t_5 \leftarrow t_5 + 1$
\ENDWHILE
\\ \textbf{3. Output optimal ${\pmb{P}^E}^*$, ${\pmb{\Theta}^E}^*$, $\pmb{f}^*$, $\{ \pmb{\alpha}_k \}^*$ $\{ \pmb{p}^I_k \}^*$ and ${\pmb{\Theta}^I}^*$}
\STATE ${\pmb{P}^E}^* \leftarrow {\pmb{P}^E}^{(t_5)}$, ${\pmb{\Theta}^E}^* \leftarrow {\pmb{\Theta}^E}^{(t_5)}$, $\pmb{f}^*  \leftarrow \pmb{f}^{(t_5)} $, $\{ \pmb{\alpha}_k \}^* \leftarrow \{ \pmb{\alpha}_k \}^{(t_5)}$, $\{ \pmb{p}^I_k \}^* \leftarrow \{ \pmb{p}^I_k \}^{(t_5)}$ and ${\pmb{\Theta}^I}^* \leftarrow {\pmb{\Theta}^I}^{(t_5)}$
\end{algorithmic}
\label{alg:solution_Problem_P}
\end{algorithm}

\section{Numerical Results}

\begin{figure}[h!]\center
\includegraphics[width= 0.5 \textwidth]{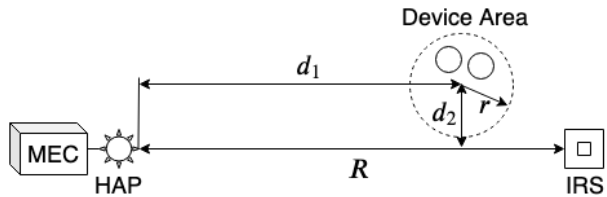}
\caption{An illustration of the locations of the HAP, of devices and of the IRS in the IRS-aided WP-MEC system.}
\label{fig:top_view}
%\hrulefill
\end{figure}

In this section, we present the pertinent numerical results, for evaluating the performance of our proposed IRS-aided WP-MEC design. %Specifically, we consider a single-cell WP-MEC network, where an $N$-element IRS is deployed for assisting the WET of and the computation offloading of $K$ single-antenna wireless devices in an OFDM-based system associated with $M$ sub-bands.
A top view of the HAP, of the wireless devices and of the IRS are shown in Fig.~\ref{fig:top_view}, where the HAP's coverage radius is $R$ and the IRS is deployed at the cell edge. The locations of wireless devices are assumed to obey the uniform distribution within a circle, whose radius and locations are specified by $r$ as well as $d_1$ and $d_2$, respectively. Their default settings are specified in the block of ``System model" in Table~\ref{tbl:simulation_parameters}. The efficiency of the energy harvesting $\eta$ is set as $0.5$. As for the communications channel, we consider both the small-scale fading and the large-scale path loss. More explicitly, the small-scale fading is assumed to be independent and identically distributed (i.i.d.) and obey the complex Gaussian distribution associated with zero mean and unit variance, while the path loss in $\rm{dB}$ is given by
\begin{IEEEeqnarray}{rCl}
\text{PL} = \text{PL}_0 - 10 \beta \log_{10} \big( \frac{d}{d_0} \big),
\end{IEEEeqnarray}
where $\text{PL}_0$ is the path loss at the reference distance $d_0$; $\beta$ and $d$ denote the path loss exponent of and the distance of the communication link, respectively. Here we use $\beta_{ua}$, $\beta_{ui}$ and $\beta_{ia}$ to represent the path loss exponent of the links between the wireless devices and the HAP, between the wireless devices and the IRS, as well as between the IRS and the HAP, respectively\footnote{We assume that the channel of the direct link between the HAP and devices is hostile (due to an obstruction), while this obstruction can be partially avoided by the IRS-reflection link. Hence, we set a higher value for $\beta_{ua}$.}. Furthermore, the additive while Gaussian noise associated with zero mean and the variable of $\sigma^2$ is imposed both on the energy signals and on the offloading signals. The default values of the parameters are set in the block of ``Communications model" in Table~\ref{tbl:simulation_parameters}. As for the computing model, the variables of $L_k$ and $c_k$ are assumed to obey the uniform distribution. The offloaded tasks are assumed to be computed in parallel by a large number of CPUs at the edge computing node, where the computing capability of each CPU is $f_{e} = 10^{9}~\rm{cycle/s}$. Then, the energy consumption at the edge for processing the offloaded computational tasks can be calculated as $\vartheta = c \kappa f_{e}^2 = 5\times 10^{-8} ~\rm{Joule/bit}$.

\begin{table}[h!]\small
\begin{center}
\caption{Default simulation parameter setting}
\label{tbl:simulation_parameters}
\begin{tabular}{ | l | r |}
\hline
Description & Parameter and Value \\ \hline \hline
\multirow{2}{*}{System model \cite{yang2019intelligent}}
& $M = 16$, $N = 30$, $K = 3$, $T=10~\rm{ms}$ \\
& $R = 12~\rm{m}$, $d_1 = 11~\rm{m}$, $d_2 = 1~\rm{m}$, $r=1~\rm{m}$ \\ \hline
Wireless energy transfer model
& $\eta = 0.5$ \\ \hline
\multirow{4}{*}{Communication model \cite{yang2019irs}}
& $B = 312.5~\rm{KHz}$ \\
& $\text{PL}_0 = 30~\rm{dB}$, $d_0 = 1~\rm{m}$, $\beta_{ua} = 3.5$, $\beta_{ui} = 2.2$, $\beta_{ia} = 2.2$  \\
& $L^d_k = 4$, $L_1 = 2$, $L_{2,k }= 3$ \\
& $\sigma^2 = 1.24\times 10^{-12}~\rm{mW}$, $\Gamma = 2$ \\ \hline
%& $p_c = 0~\rm{mW}$ \\ \hline
\multirow{3}{*}{Computing model \cite{wang2018joint}}
& $L_k = [15, 20]~\rm{Kbit}$ \\
& $c_k = [400,500]~\rm{cycle/bit}$ \\
& $f_{max} = 1 \times 10^8~\rm{cycle/s}$ \\
& $\kappa = 10^{-28}$, $\vartheta = 5\times 10^{-8}~\rm{Joule/bit}$ \\ \hline
Convergence criterion &  $\epsilon = 0.001$ \\ \hline
\end{tabular}
\end{center}
\end{table}

Apart from our algorithms developed in Section~\ref{sec:solution}, we also consider two benchmark schemes for comparison. Let us describe these three schemes as follows.
\begin{itemize}
\item \emph{With IRS}: In this scheme, we optimize both the power allocation $\pmb{p}^E$ and the IRS reflection coefficients $\pmb{\Theta}^E$ at the WET phase, as well as the local CPU frequency at devices $\pmb{f}$, the sub-band-device association $\{\pmb{\alpha}_k\}$, the power allocation $\{\pmb{p}_k\}$ and the IRS reflection coefficients $\pmb{\Theta}^I$ at the computing phase, relying on Algorithm~\ref{alg:solution_Problem_P}.
\item \emph{RandPhase}: The power allocation $\pmb{p}^E$ at the WET phase, as well as the local CPU frequency at devices $\pmb{f}$, the sub-band-device association $\{\pmb{\alpha}_k\}$ and the power allocation $\{\pmb{p}_k\}$ at the computing phase are optimized with the aid of Algorithm~\ref{alg:solution_Problem_P}, while we skip the design of the IRS reflection coefficients $\pmb{\Theta}^E$ and $\pmb{\Theta}^I$, whose amplitude response is set to $1$ and phase shifts are randomly set in the range of $[0,2\pi)$ obeying the uniform distribution.
\item \emph{Without IRS}: The composite channel $\pmb{f}^H_m \pmb{V}_k \pmb{\Theta}$ is set to $0$ both for the WET and for the computation offloading. The power allocation $\pmb{p}^E$ at the WET phase, as well as the local CPU frequency at devices $\pmb{f}$, the sub-band-device association $\{\pmb{\alpha}_k\}$ and the power allocation $\{\pmb{p}_k\}$ at the computing phase are optimized with the aid of Algorithm~\ref{alg:solution_Problem_P}, while we skip the optimization of the IRS reflection coefficient $\pmb{\Theta}^E$ and $\pmb{\Theta}^I$.
\end{itemize}

Let us continue by presenting the selection of the time allocation, sub-band allocation in the WET and the computing phases, as well as the impact of diverse environment settings, as follows.

\subsection{Selection of the Time Allocation}
\label{sec:Selection of the Time Allocation}

\begin{figure}[h!]\center
\includegraphics[width=3.2in]{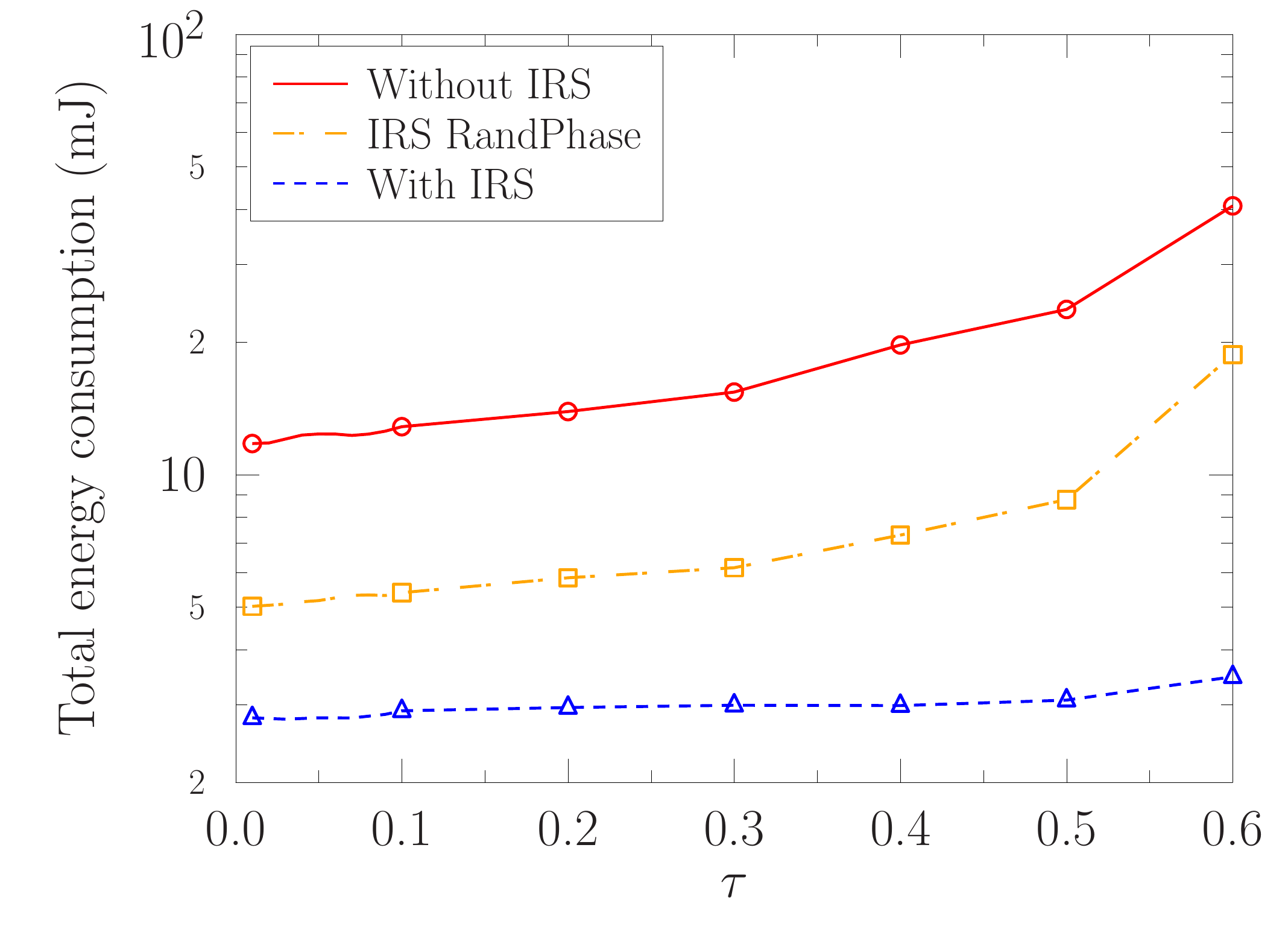}
\caption{Simulation results of the total energy consumption versus the time allocation $\tau$. The parameter settings are specified in Table~\ref{tbl:simulation_parameters}.}
\label{fig:selection_of_tau}
%\hrulefill
\end{figure}

In order to find an appropriate time allocation for our WP-MEC system, we depict the total energy consumption (the OF of Problem~$\mathcal{P}1$) versus the the time allocation $\tau$ in Fig.~\ref{fig:selection_of_tau}. It can be seen that the total energy consumption becomes higher upon increasing $\tau$ for all these three schemes considered. The reason behind it is explained as follows. For a given volume of the computational task to be offloaded within the time duration of $T$, an increase of $\tau$ implies a higher offloading rate required by computation offloading, while
at a glance of \eqref{eqn:achievable_rate}, the computation offloading rate is formulated as a logarithmic function of the offloading power. Hence, we have to largely increase the transmit power of computation offloading for providing the extra offloading rate required by the increase of $\tau$, which results in a higher energy consumption at the wireless devices. Furthermore, since the energy required by WET is determined by the energy consumption at the wireless devices, the total energy consumption becomes higher upon increasing $\tau$.
Based on this discussion, it seems that we should select the value of $\tau$ as small as possible. However, this may lead to an upsurge of the power consumption for WET, which might exceed the maximum allowable transmit power at the HAP. Therefore, as a compromise, for the environment associated with the default settings we select $\tau = 0.1$, beyond which the total energy consumption becomes increasingly higher along with $\tau$.

\subsection{Joint Sub-Band and Power Allocation in the WET and Computing Phases}

\begin{figure*}[h!]\center
\subfloat[]{\includegraphics[width=3.2in]{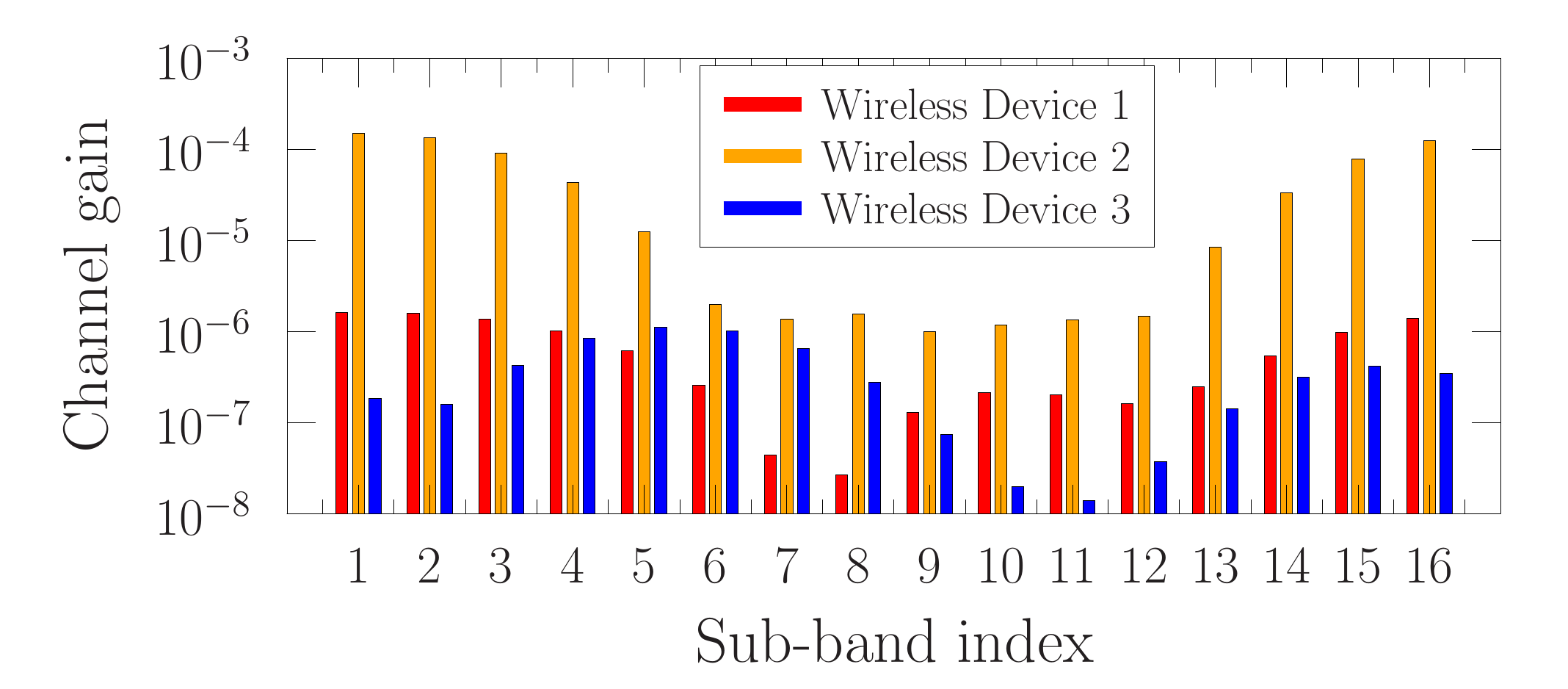}\label{fig:channel_WET}}
\subfloat[]{\includegraphics[width=3.2in]{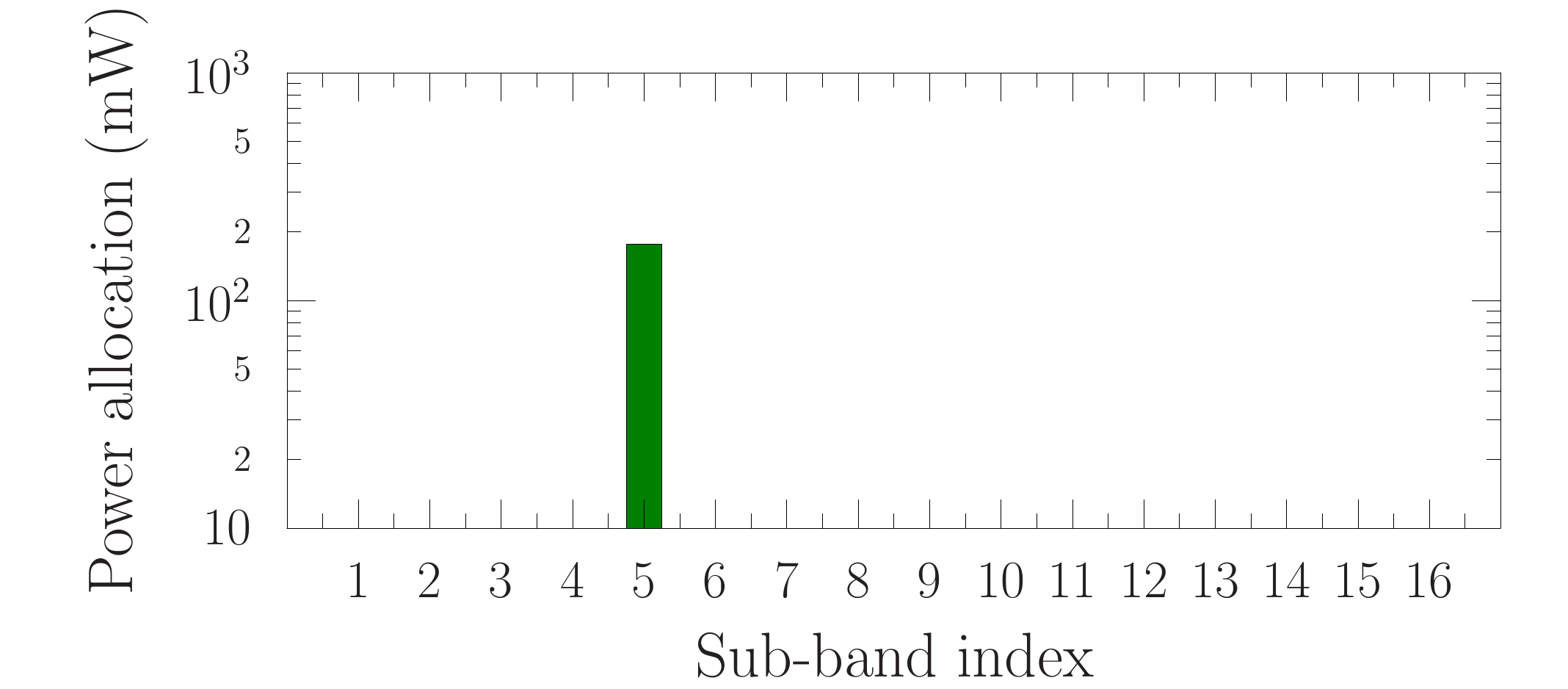}\label{fig:p_E_WET}} \\
\subfloat[]{\includegraphics[width=3.2in]{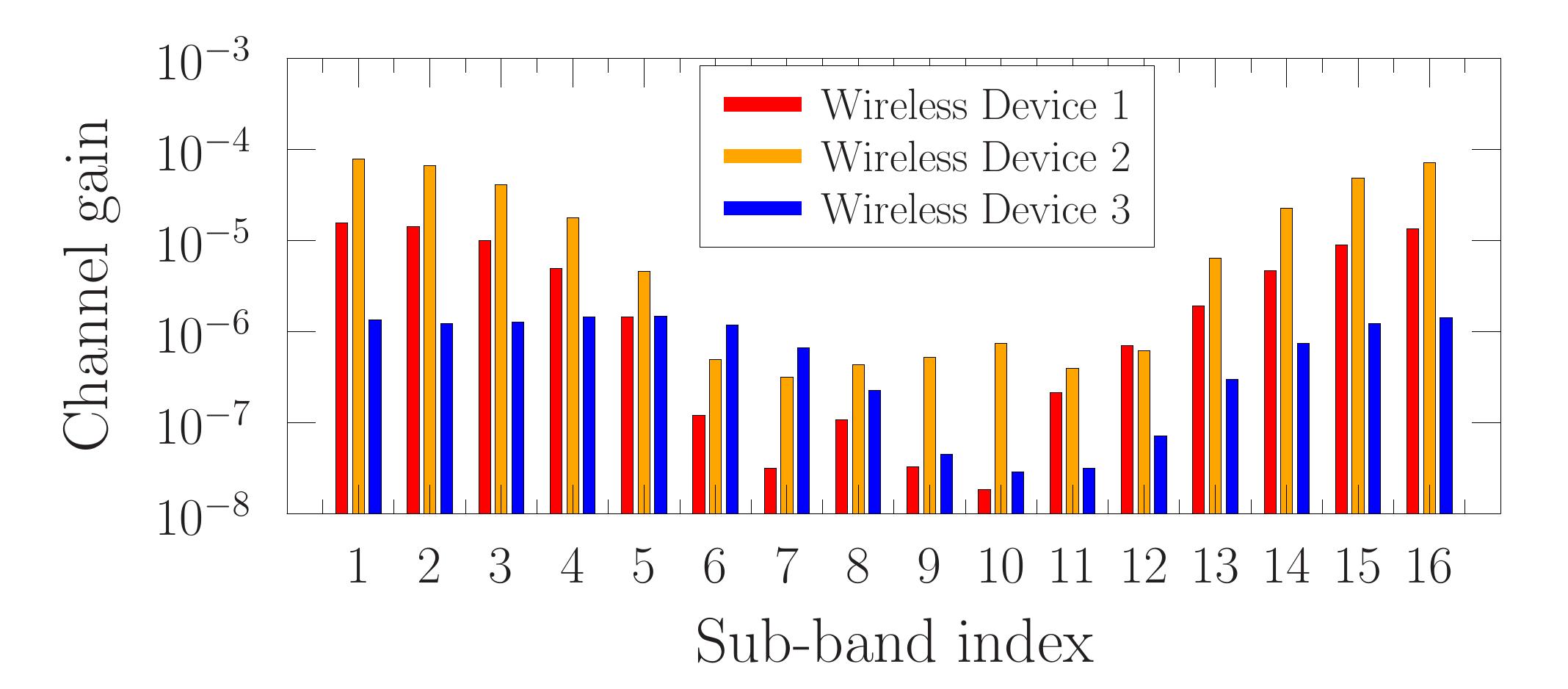}\label{fig:channel_computing}}
\subfloat[]{\includegraphics[width=3.2in]{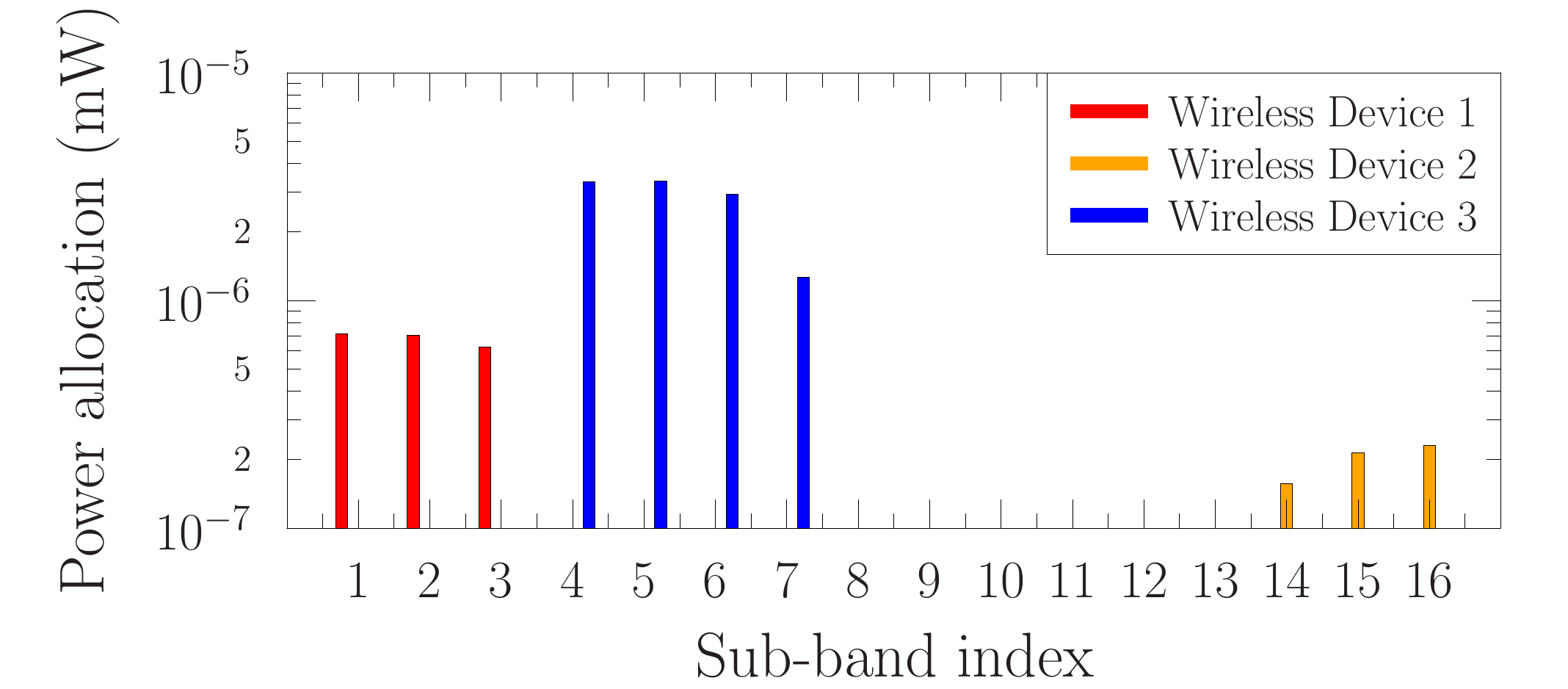}\label{fig:p_E_computing}}
\caption{Joint sub-band and of power allocation for the WET and the computing phases, relying on the Algorithm~\ref{alg:solution_Problem_P}, where the number of bits to be processed is set the same as $20~\rm{Kbits}$ for the three wireless devices.
(a) The channel gain at the WET phase; (b) The joint sub-band and power allocation at the WET phase; (3) The channel gain at the computing phase; (d) The joint sub-band and power allocation at the computing phase. The parameter settings are specified in Table~\ref{tbl:simulation_parameters}.}
\label{fig:allocation}
\end{figure*}

Fig.~\ref{fig:allocation} illustrates the channel gain as well as the joint sub-band and power allocation both for the WET and computing phases. Our observations are as follows. Firstly, as shown in Fig.~\ref{fig:p_E_WET}, only the $5$-th sub-band is activated for WET. This allocation is jointly determined by the power consumption of the computing phase and by the channel gain in the WET phase. Specifically, with the reference of Fig.~\ref{fig:p_E_computing}, Device 3 requires the highest power consumption for computation offloading. Given that the overall performance is dominated by the device having the highest energy consumption, we may reduce the energy consumption of WET, by activating the sub-band associated with the highest channel gain of Device 3, which is the $5$-th sub-band as shown in Fig.~\ref{fig:channel_WET}.
Secondly, with the reference of Fig.~\ref{fig:channel_computing}, it can be observed that the power allocation in Fig.~\ref{fig:p_E_computing} obeys the water-filling principle for each device, i.e. allocating a higher power to the sub-band possessing a high channel gain. This corresponds to the power allocation obtained in \eqref{eqn:p_k_m}.
Thirdly, comparing Fig.~\ref{fig:channel_WET} and Fig.~\ref{fig:channel_computing}, we can see that the channel gains in the WET and computing phase are different for each device after we optimize the IRS reflection coefficients, which consolidates our motivation to conceive separate IRS designs for the WET and the computing phases.

\subsection{Performance of the Proposed Algorithms}

In order to evaluate the benefits of employing an IRS in WP-MEC systems, we compare the performance of our proposed algorithms with that of the benchmark schemes, under various settings of the number of IRS reflection elements, of the device location, of the path loss exponent of the IRS-related channel, and of the energy consumption per bit at the edge, as follows.

\subsubsection{Impact of the Number of IRS Reflection Elements}

\begin{figure}[h!]\center
\centering
\begin{minipage}{.47\textwidth}
  \centering
  \includegraphics[width=1\textwidth]{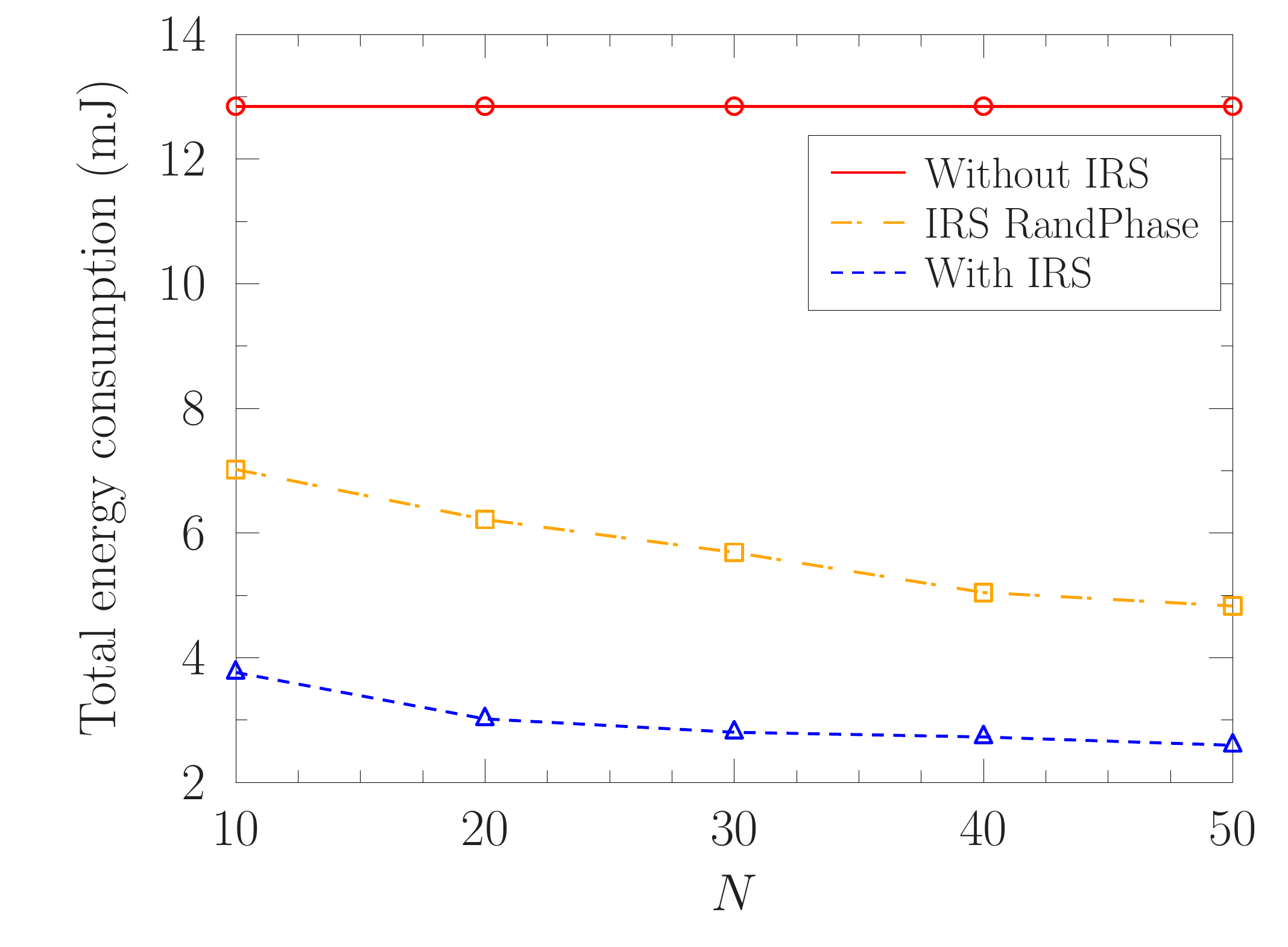}
  \captionof{figure}{Simulation results of the total energy consumption versus the number of IRS reflection elements $N$. The rest of parameters are specified in Table~\ref{tbl:simulation_parameters}.}
  \label{fig:N}
\end{minipage}%
\hspace{0.2cm}
\begin{minipage}{.47\textwidth}
  \centering
  \includegraphics[width=1\textwidth]{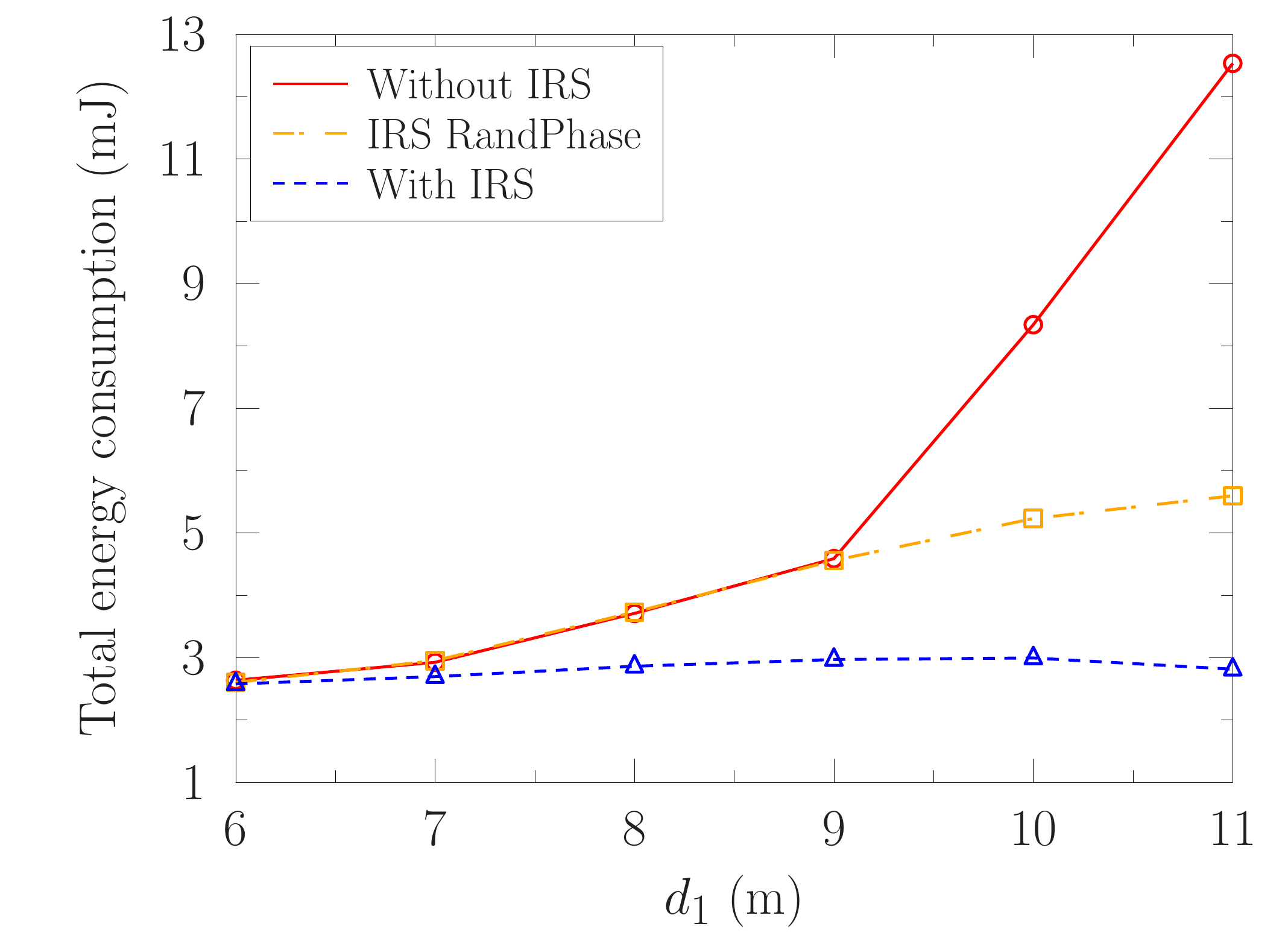}
  \captionof{figure}{Simulation results of the total energy consumption versus the distance between the HAP and the wireless device circle $d_1$. Other parameters are set in Table~\ref{tbl:simulation_parameters}.}
  \label{fig:d_1}
\end{minipage}
\end{figure}

%\begin{figure}[h!]\center
%\includegraphics[width=3.2in]{3_Fig/N.eps}
%\caption{Simulation results of the total energy consumption versus the number of IRS reflection elements $N$. The rest of parameters are specified in Table~\ref{tbl:simulation_parameters}.}
%\label{fig:N}
%\end{figure}

Fig.~\ref{fig:N} shows the simulation results of the total energy consumption versus the number of IRS reflection elements for the three schemes considered. We have the following observations.
Firstly, the performance gap between the scheme ``Without IRS" and the scheme ``IRS RandPhase" increases along with $N$, which implies that the IRS is capable of assisting the energy consumption reduction in the WP-MEC system, even without carefully designing the IRS reflection coefficients. This is due to the so-called virtual array gain induced by the IRS, as mentioned in Section~\ref{sec:Introduction}.
Secondly, the scheme ``With IRS" outperforms the scheme ``IRS RandPhase", which indicates that our sophisticated design of IRS reflection coefficients may provide the so-called passive beamforming gain for computation offloading.
Note that different from the conventional MEC systems \cite{bai2019latency} where WET is not employed, these two types of gain are exploited twice in WP-MEC systems (during the WET and computing phases, respectively).
As such, IRSs are capable of efficiently reducing the energy consumption in WP-MEC systems.

\subsubsection{Impact of the Distance between the Device Circle and the IRS}

%\begin{figure}[h!]\center
%\includegraphics[width=3.2in]{3_Fig/d_1.eps}
%\caption{Simulation results of the total energy consumption versus the distance between the HAP and the wireless device circle $d_1$. Other parameters are set in Table~\ref{tbl:simulation_parameters}.}
%\label{fig:d_1}
%\end{figure}

Fig.~\ref{fig:d_1} presents the simulation results of the total energy consumption versus the distance between the HAP and the mobile wireless circles. Our observations are as follows.
Firstly, the two IRS-aided schemes do not show any visible advantage over the scheme of ``Without IRS" when we have $d_1 < 6~\rm{m}$, which indicates that each IRS has a limited coverage.
Secondly, the benefit of deploying the IRS is becomes visible at $d_1 > 9~\rm{m}$ in the scheme of ``IRS RandPhase", while the advantage of the ``With IRS" scheme is already notable at $d_1 = 7~\rm{m}$. This observation implies that our sophisticated design of IRS reflection coefficient is capable of extending the coverage of IRS.

\begin{figure}[h!]\center
\centering
\begin{minipage}{.47\textwidth}
  \centering
  \includegraphics[width=1\textwidth]{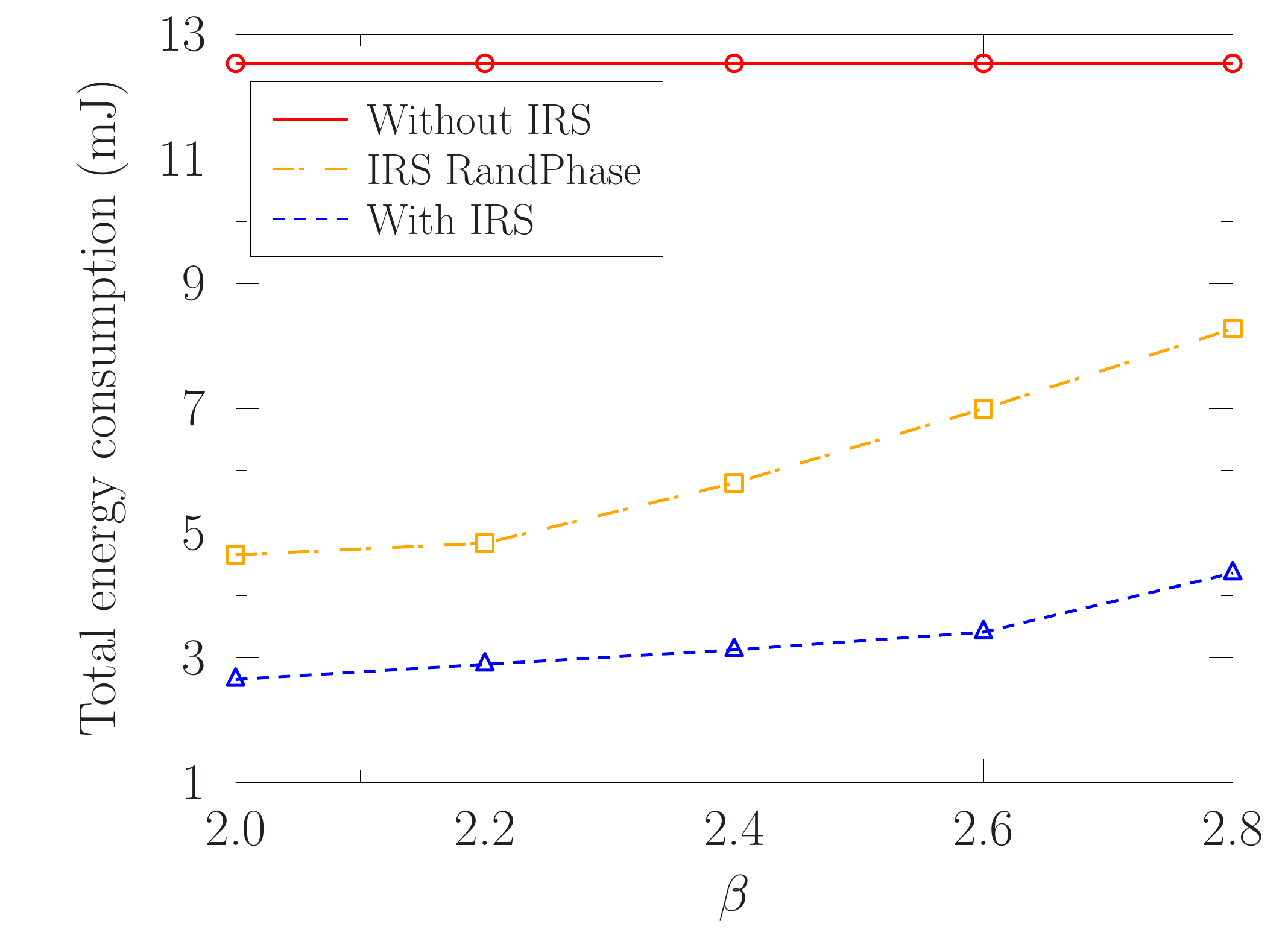}
  \captionof{figure}{Simulation results of the total energy consumption versus the path loss exponent of the IRS reflection link $\beta$, where we set $\beta_{ui} = \beta_{ia} = \beta$. Other parameters are set in Table~\ref{tbl:simulation_parameters}.}
  \label{fig:beta}
\end{minipage}%
\hspace{0.2cm}
\begin{minipage}{.47\textwidth}
  \centering
  \includegraphics[width=1\textwidth]{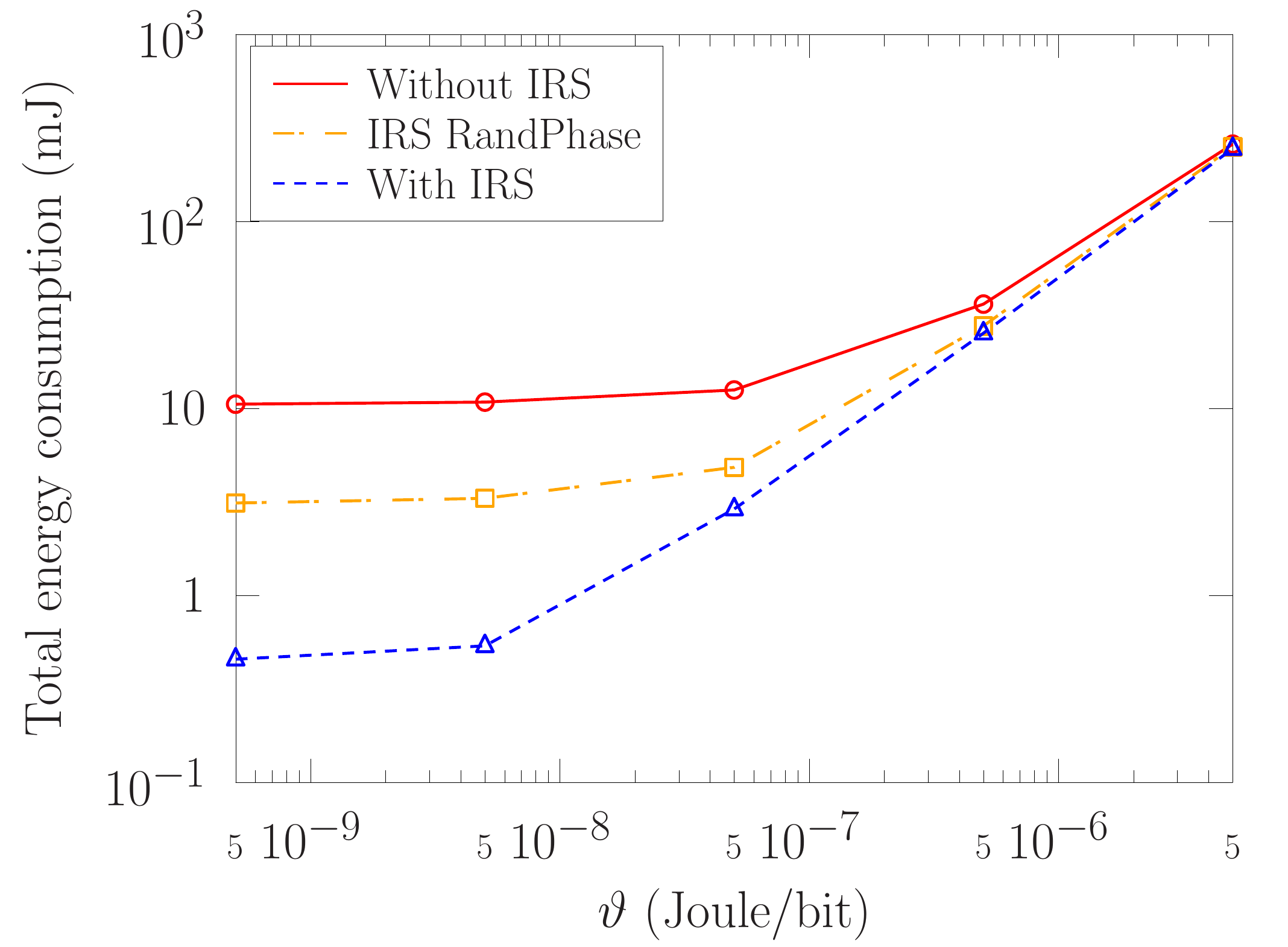}
  \captionof{figure}{Simulation results of the total energy consumption versus the energy consumption per bit at the edge. Other parameters are set in Table~\ref{tbl:simulation_parameters}.}
  \label{fig:vartheta}
\end{minipage}
\end{figure}

\subsubsection{Impact of Path Loss Exponent}

%\begin{figure}[h!]\center
%\includegraphics[width=3.2in]{3_Fig/beta.eps}
%\caption{Simulation results of the total energy consumption versus the path loss exponent of the IRS reflection link $\beta$, where we set $\beta_{ui} = \beta_{ia} = \beta$. Other parameters are set in Table~\ref{tbl:simulation_parameters}.}
%\label{fig:beta}
%\end{figure}

Fig.~\ref{fig:beta} depicts the simulation results of the total energy consumption versus the path loss exponent of the IRS related links. It can be seen that the total energy consumption decreases if a higher path loss exponent is encountered, which is because a higher $\beta$ leads to a lower channel gain of the IRS-reflected link. This observation provides an important engineering insight: the locations of IRSs should be carefully selected for avoiding obstacles.

\subsubsection{Impact of energy consumption at the edge}

%\begin{figure}[h!]\center
%\includegraphics[width=3.2in]{3_Fig/vartheta.eps}
%\caption{Simulation results of the total energy consumption versus the energy consumption per bit at the edge. Other parameters are set in Table~\ref{tbl:simulation_parameters}.}
%\label{fig:vartheta}
%\end{figure}

Fig.~\ref{fig:vartheta} shows the simulation results of the total energy consumption versus the energy consumption per bit at the edge node. It can be observed that the advantage of deploying IRS is eminent when we have a small value of $\vartheta$, while the benefit becomes smaller upon increasing the value of $\vartheta$. The reason is explained as follows. The OF of Problem~$\mathcal{P}1$ is the combination of the energy consumption of WET and of processing the offloaded computational tasks. If the energy consumption per bit at the edge node is of a small value, the energy consumption of WET plays a dominant role in the total energy consumption. In this case, the benefit of employing IRS is significant. By contrast, if $\vartheta$ becomes higher, the total energy consumption is dominated by that at the edge. In this case, although the energy consumption of WET can be degraded by deploying IRSs, this reduction becomes marginal.

\section{Conclusions}

To reduce the energy consumption of WP-MEC systems, we have proposed an IRS-aided WP-MEC scheme and formulate an energy minimization problem. A sophisticated algorithm has been developed for optimizing the settings both in the WET and the computing phases. Our numerical results reveal the following insights.
Firstly, the employment of IRSs is capable of substantially reducing the energy consumption of the WP-MEC system, especially when the IRS is deployed in vicinity of wireless devices.
Secondly, the energy consumption decreases upon increasing the number of IRS reflection elements.
Thirdly, the locations of IRSs should be carefully selected for avoiding obstacles.
These results inspire us to conceive a computational rate maximization design for the IRS-aided WP-MEC system as a future work.

\bibliographystyle{ieeetr}
%\bibliography{mybib.bib}
\bibliography{IEEEabrv.bib}

% that's all folks
\end{document}